\documentclass{jfm}

\usepackage{graphicx}
\usepackage{natbib}

\ifCUPmtlplainloaded \else
  \checkfont{eurm10}
  \iffontfound
    \IfFileExists{upmath.sty}
      {\typeout{^^JFound AMS Euler Roman fonts on the system,
                   using the 'upmath' package.^^J}%
       \usepackage{upmath}}
      {\typeout{^^JFound AMS Euler Roman fonts on the system, but you
                   dont seem to have the}%
       \typeout{'upmath' package installed. JFM.cls can take advantage
                 of these fonts,^^Jif you use 'upmath' package.^^J}%
      }
  \else
  \fi
\fi

\ifCUPmtlplainloaded \else
  \checkfont{msam10}
  \iffontfound
    \IfFileExists{amssymb.sty}
      {\typeout{^^JFound AMS Symbol fonts on the system, using the
                'amssymb' package.^^J}%
       \usepackage{amssymb}%

      }{}
  \fi
\fi

\ifCUPmtlplainloaded \else
  \IfFileExists{amsbsy.sty}
    {\typeout{^^JFound the 'amsbsy' package on the system, using it.^^J}%
     \usepackage{amsbsy}}
    {\providecommand\boldsymbol[1]{\mbox{\boldmath $##1$}}}
\fi

\newsavebox{\astrutbox}
\sbox{\astrutbox}{\rule[-5pt]{0pt}{20pt}}

\title[Regular shock refraction]{An exact Riemann solver based solution for regular shock refraction}

\author[Delmont, Keppens and van der Holst]%
{P.\ns D\ls E\ls L\ls M\ls O\ls N\ls T\ls$^{1,2}$%
  \thanks{Email: Peter.Delmont@wis.kuleuven.be},\ns
R.\ns K\ls E\ls P\ls P\ls E\ls N\ls S\ls$^{1,2,3,4}$ \break
\and B.\ns v\ls a\ls n\ns d\ls e\ls r\ns  H\ls O\ls L\ls S\ls T\ls$^5$}

\affiliation{
$^1$ Centre for Plasma Astrophysics, K.U.Leuven, Heverlee, Belgium\\
$^2$ Leuven Mathematical Modeling and Computational Science Centre, Heverlee, Belgium\\
$^3$ Astronomical Institute, Utrecht University, The Netherlands\\
$^4$ FOM institute for Plasma Physics Rijnhuizen, Nieuwegein, The Netherlands\\
$^5$ Centre for Space Environment Modeling, Ann Arbor MI, USA
}
\pubyear{??}
\volume{??}
\pagerange{??}
\date{\today}

\begin{document}

\maketitle

\begin{abstract}
We study the classical problem of planar shock refraction at an oblique density discontinuity, separating two gases at rest.  When the shock impinges on the density discontinuity, it refracts and in the hydrodynamical case $3$ signals arise. Regular refraction means that these signals meet at a single point, called the \textit{triple point}.

 After reflection from the top wall, the contact discontinuity becomes unstable due to local Kelvin-Helmholtz instability, causing the contact surface to roll up and develop the Richtmyer-Meshkov instability.  We present an exact Riemann solver based solution strategy to describe the initial self similar refraction phase, by which we can  quantify the vorticity deposited on the contact interface.  We investigate the effect of a perpendicular magnetic field and quantify how addition of a perpendicular magnetic field increases the deposition of vorticity on the contact interface slightly under constant \textit{Atwood Number}.  We predict wave pattern transitions, in agreement with experiments, von Neumann shock refraction theory, and numerical simulations performed with the grid-adaptive code AMRVAC.  These simulations also describe the later phase of the Richtmyer-Meshkov instability.
\end{abstract}

\section{Introduction}

We study the classical problem of regular refraction of a shock at an oblique density discontinuity.  Long ago, \cite{NE43} deduced the critical angles for regularity of the refraction, while \cite{TA47} found relations between the angles of refraction.  Later on, \cite{HE66} extended this work to irregular refraction by use of polar diagrams.  An example of an early shock tube experiment was performed by \cite{JA56}.  Amongst many others, \cite{AH78a, AH78b} performed experiments in which also irregular refraction occured.

In 1960, Richtmyer performed the linear stability analysis of the interaction of shock waves with density discontinuities, and concluded that the shock-accelerated contact is unstable to perturbations of all wavelenghts, for \textit{fast-slow} interfaces (\cite{RI60}).  In hydrodynamics (HD) an interface is said to be fast-slow if $\eta > 1$, and \textit{slow-fast} otherwise, where $\eta$ is the density ratio across the interface (figure~\ref{fig:setup}).  The instability is not a classical fluid instability in the sense that the perturbations grow linearly and not exponentially.  The first experimental validation was performed by \cite{ME69}.  On the other hand, according to linear analysis the interface remains stable for slow-fast interfaces.  This misleading result is only valid in the linear phase of the process and near the triple point: a wide range of experimental (e.g. \cite{AH78b}) and numerical (e.g. \cite{NO05}) results show that also in this case the interface becomes unstable.  The growth rates obtained by linear theory compare poorly to experimentally determined growth rates (\cite{ST87}).  The governing instability is referred to as the Richtmyer-Meshkov instability (RMI) and is nowadays a topic of research in e.g. inertial confinement fusion ( e.g. \cite{OR99}), astrophysics (e.g. \cite{KI06}), and it is a common test problem for numerical codes ( e.g. \cite{HO07}).

In essence, the RMI is a local Kelvin-Helmholtz instability, due to the deposition of vorticity on the shocked contact.  \cite{HZ89} formulate an interesting vortex paradigm, which describes the process of shock refraction, using vorticity as a central concept.  Later on, \cite{SA98} performed an extensive analysis of the baroclinic circulation generation on shocked slow-fast interfaces.

A wide range of fields where the RMI occurs, involves ionized, quasi-neutral plasmas, where the magnetic field plays an important role.  Therefore, more recently there has been some research done on the RMI in magnetohydrodynamics (MHD).  \cite{SA03} proved by numerical simulations, exploiting Adaptive Mesh Refinement (AMR), that the RMI is suppressed in planar MHD, when the initial magnetic field is normal to the shock.  \cite{WH05} solved the problem of planar shock refraction analytically, making initial guesses for the refracted angles.  The basic idea is that ideal MHD does not allow for a jump in tangential velocity, if the magnetic field component normal to the contact discontinuity (CD), does not vanish (see e.g. \cite{GP04}).  The solution of the Riemann problem in ideal MHD is well-studied in the literature (e.g. \cite{LA57}), and due to the existence of three (slow, Alfv\'en, fast) wave signals instead of one (sound) signal, it is much richer than the HD case.  The Riemann problem usually considers the self similar temporal evolution of an initial discontinuity, while we will consider stationary two dimensional conditions.  The interaction of small perturbations with MHD (switch-on and switch-off) shocks was studied both analytically by \cite{TO66} and numerically by \cite{CT67}.  Later on, the evolutionarity of intermediate shocks, which cross the Alfv\'en speed, has been studied extensively.  Intermediate shocks are unstable under small perturbations, and are thus not evolutionary.  \cite{BW88} and \cite{ST98} found intermediate shocks in respectively one and two dimensional simulations.  The evolutionary condition became controversial and amongst others \cite{MR97a,MR97b} argue that the evolutionary condition is not relevant in dissipative MHD.  \cite{CH93} reported a $2 \rightarrow 4$ intermediate shock observed by \textit{Voyager 1} in 1980 and \cite{FW08} recognised a $2 \rightarrow 3$ intermediate shock, which was observed by \textit{Voyager 2} in 1979.  On the other hand, \cite{BA96} argue that if the full set of MHD equations is used to solve planar MHD, a small tangential disturbance on the magnetic field vector splits the rotational jump from the compound wave, transforming it into a slow shock.  They investigate the reconstruction process of the non-evolutionary compound wave into evolutionary shocks.  Also \cite{FK97,FK01} do not reject the evolutionary condition, and develop a shock capturing scheme for evolutionary solutions in MHD,  However, since all the signals in this paper are essentially hydrodynamical, we do not have to worry about evolutionarity for the setup considered here. 

In this paper, we solve the problem of regular shock refraction exactly, by developing a stationary two-dimensional Riemann solver.  Since a normal component of the magnetic field suppresses the RMI, we investigate the effect of a perpendicular magnetic field.  The transition from slow-fast to fast-slow refraction is described in a natural way and the method can predict wave pattern transitions.  We also perform numerical simulations using the grid-adaptive code AMRVAC (\cite{HO07,KE03}).

In section 2, we formulate the problem and introduce the governing MHD equations.  In section 3, we present our Riemann solver based solution strategy and in section 4, more details on the numerical implementation are described.  Finally, in section 5, we present our results, including a case study, the prediction of wave pattern transitions, comparison to experiments and numerical simulations, and the effect of a perpendicular magnetic field on the stability of the CD.

\section{Configuration and governing equations}

\subsection{Problem setup}

\begin{figure}
\centering
\includegraphics[width=.99\textwidth,viewport= 145 690 525 800,clip]{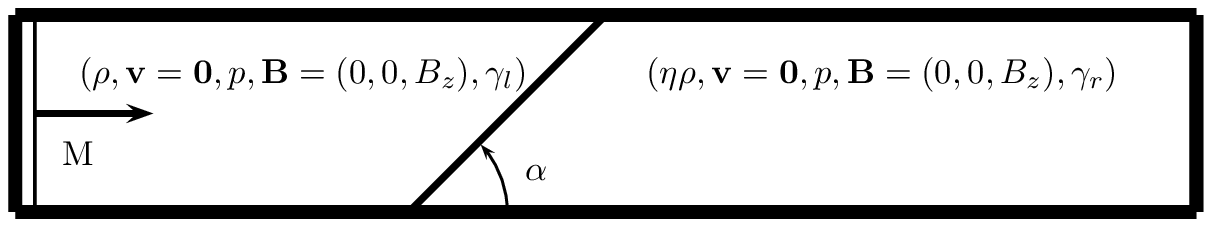}
\caption{ Initial configuration:  a shock moves with shock speed $M$ to an inclined density discontinuity.  Both the upper and lower boundary are solid walls, while the left and the right boundaries are open.}
\label{fig:setup}
\end{figure}


%

As indicated in figure~\ref{fig:setup}, the hydrodynamical problem of regular shock refraction is parametrised by $5$ independent initial parameters:  the angle $\alpha$ between the shock normal and the initial density discontinuity CD, the sonic Mach number $M$ of the impinging shock, the density ratio $\eta$ across the $CD$ and the ratios of specific heat $\gamma_l$ and $\gamma_r$ on both sides of the CD.  The shock refracts in $3$ signals: a reflected signal (R), a transmitted signal (T) and a shocked contact discontinuity (CD), where we allow both R and T to be expansion fans or shocks.
Adding a perpendicular magnetic field, $B$, also introduces the plasma-$\beta$ in the pre-shock region,\begin{equation} \beta = \frac{2p}{B^2}, \label{eq:beta} \end{equation} which is in our setup a sixth independent parameter.  As argued later, the shock then still refracts in $3$ signals (see figure~\ref{fig:states}): a reflected signal (R), a transmitted signal (T) and a shocked contact discontinuity (CD), where we allow both $R$ and $T$ to be expansion fans or shocks.

\subsection{Stationary MHD equations}

In order to describe the dynamical behaviour of ionized, quasi-neutral plasmas, we use the framework of ideal MHD.  We thereby neglect viscosity and resistivity, and suppose that the length scales of interest are much larger than the Debye length and there are enough particles in a Debye sphere (see e.g. \cite{GP04}).  As written out in conservative form and for our planar problem, the stationary MHD equations are

\begin{equation}
\frac{\partial}{\partial x}\mathbf{F}+\frac{\partial}{\partial y}\mathbf{G} =\mathbf{0},
\label{eq:statmhd}
\end{equation}
where we introduced the flux terms
\begin{equation}\mathbf{F} = \left( \rho v_x, \rho v_x^2 + p +\frac{B^2}{2}, \rho v_x v_y, v_x (\frac{\gamma}{\gamma- 1}p+\rho \frac{v_x^2 + v_y^2}{2} + B^2), v_x B ,v_x \gamma \rho \right)^t,\end{equation} and \begin{equation}\mathbf{G} = \left(\rho v_y, \rho v_x v_y, \rho v_y^2 + p+\frac{B^2}{2}, v_y (\frac{\gamma}{\gamma-1}p+\rho \frac{v_x^2 + v_y^2}{2}+B^2),v_y B, v_y \gamma \rho \right)^t.\end{equation}
The applied magnetic field  $\mathbf{B} = \left( 0,0, B \right)$ is assumed purely perpendicular to the flow and  the velocity $\mathbf{v} = (v_x,v_y,0)$.  Note that the ratio of specific heats, $\gamma$, is interpreted as a variable, rather than as an equation parameter, which is done to treat gases and plasmas in a simple analytical and numerical way.  The latter equation of the system expresses that $\nabla \cdot (\gamma \rho \mathbf{v})=0$.  Also note that $\nabla \cdot \mathbf{B} = 0$ is trivially satisfied.  

\subsection{Planar stationary Rankine-Hugoniot condition}

\begin{figure}
\centering
\includegraphics[width=.47\textwidth,viewport= 50 620 370 880,clip]{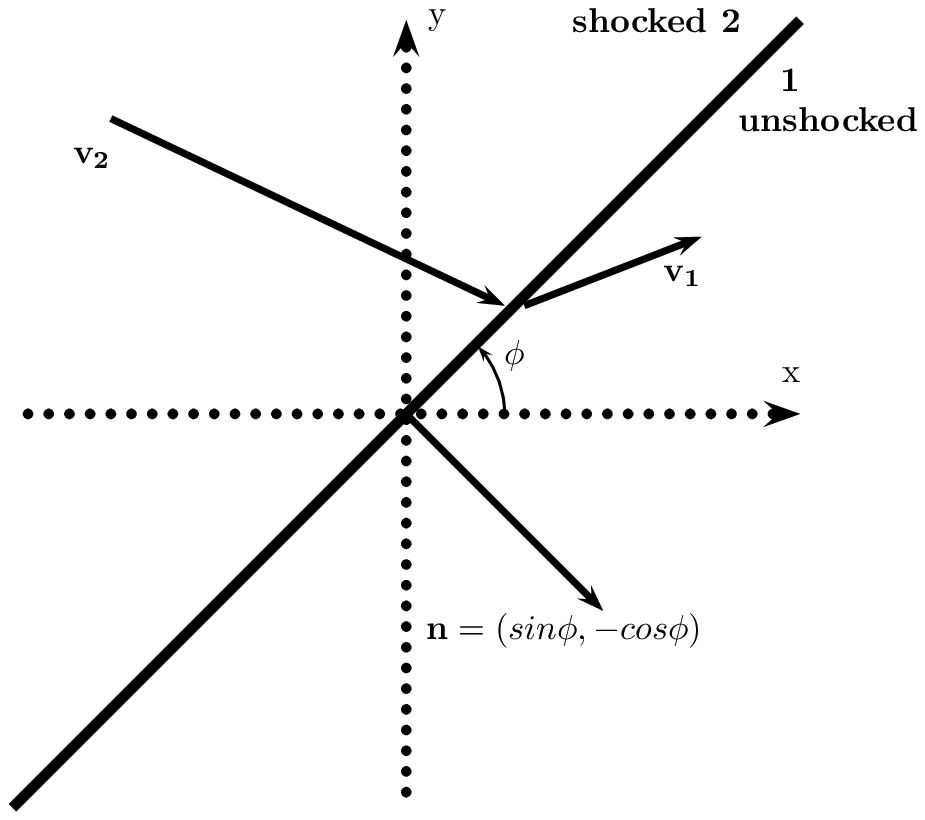}
\includegraphics[width=.47\textwidth, viewport= 00 520 370 880,clip]{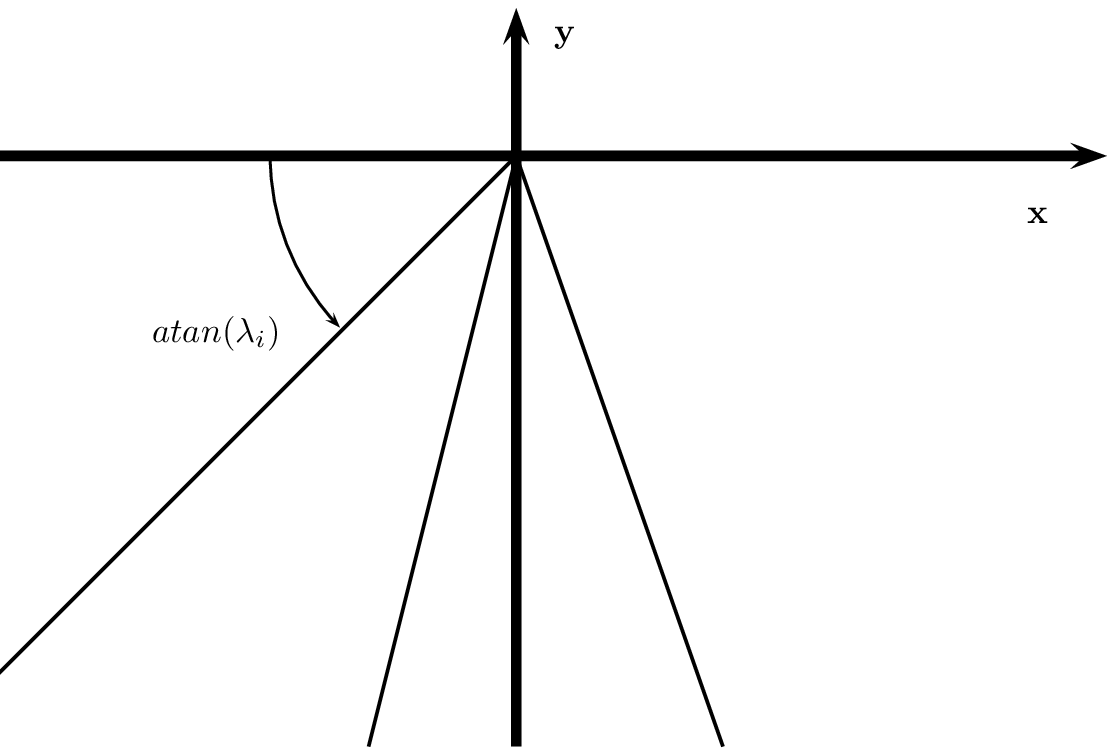}
\caption[setup]{Left: A stationary shock, seperating two constant states across an inclined planar discontinuity. Right: The eigenvalues of the matrix $\mathbf{A}$ from (\ref{eq:A}) correspond to the refracted signals.}
\label{fig:rh}
\end{figure}

We allow \textit{weak solutions} of the system, which are solutions of the integral form of the MHD equations.  The shock occuring in the problem setup, as well as those that later on may appear as $R$ or $T$ signals obey the \textit{Rankine-Hugoniot conditions}.  In the case of two dimensional stationary flows (see figure~\ref{fig:rh}), where the shock speed $s=0$, the Rankine-Hugoniot conditions follow from equation (\ref{eq:statmhd}).  When considering a thin continuous transition layer in between the two regions, with thickness $\delta$, solutions of the integral form of equation (\ref{eq:statmhd}) should satisfy $\mathop {\lim }\limits_{\delta \to 0 } \int_1^2 ( \frac{\partial}{\partial x}\mathbf{F}+\frac{\partial }{\partial y}\mathbf{G}) dl =0$.  For vanishing thickness of the transition layer this yields the Rankine-Hugoniot conditions as

\begin{eqnarray} 
- \mathop {\lim }\limits_{\delta \to 0 } \int_1^2 \left(\frac{1}{\sin \phi} \frac{\partial }{\partial l}\mathbf{F} -\frac{1}{\cos \phi} \frac{\partial }{\partial l}\mathbf{G}\right) dl &=& 0\\
& \Updownarrow  & \nonumber\\
\left[ \left[  \mathbf{F}\right] \right] & = & \xi \left[ \left[ \mathbf{G} \right] \right] ,
\label{eq:statrh}
\end{eqnarray}
where $\xi = \tan \phi$ and $\phi$ is the angle between the $x$-axis and the shock as indicated in figure~\ref{fig:rh}.  The symbol $[[\ ]]$ indicates the jump across the interface.

\section{Riemann Solver based solution strategy}

\subsection{Dimensionless representation}

\begin{figure}
\centering
\includegraphics[width=.8\textwidth,viewport= 100 670 625 900,clip]{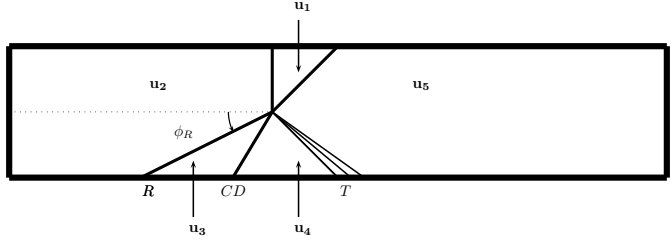}
\caption{The wave pattern during interaction of the shock with the $CD$.  The upper and lower boundaries are rigid walls, while the left and right boundaries are open.}
\label{fig:states}
\end{figure}

In this section we present how we initialise the problem in a dimensionless manner.  In the initial refraction phase, the shock wil introduce 3 wave signals (R, CD, T), and 2 new constant states develop, as schematically shown in figure~\ref{fig:states}.  We choose a representation in which the initial shock speed $s$ equals its sonic Mach number $M$.  We determine the value of the primitive variables in the post-shock region by applying the stationary Rankine-Hugoniot conditions in the shock rest frame.  In absence of a magnetic field, we use a slightly different way to nondimensionalise the problem.  Note $\mathbf{u}_i = \left( \rho_i , v_{x,i},v_{y,i}, p_{tot,i}, B_i, \gamma_i \right)$, where the index $i$ refers to the value taken in the $i-$th region (figure~\ref{fig:states}) and the total pressure
\begin{equation}
 p_{tot}=p+\frac{B^2}{2}.
\label{eq:ptot}
\end{equation}
In the $HD$ case, we define $p=1$ and $\rho = \gamma_l$ in $\mathbf{u}_1$.  Now all velocity components are scaled with respect to the sound speed in this region between the impinging shock and the initial CD.   Since this region is initially at rest, the sonic Mach number $M$ of the shock equals its shock speed $s$.  When the shock intersects the CD, the triple point follows the unshocked contact slip line.  It does so at a speed $\mathbf{v}_{tp}=(M, M \tan \alpha)$, therefore we will solve the problem in the frame of the stationary triple point.  We will look for selfsimilar solutions in this frame, $\mathbf{u}=\mathbf{u}(\phi)$, where all signals are stationary.  We now have that $\tilde{v}_x=v_x-M$ and $\tilde{v}_y=v_y-M \tan \alpha$, where $\tilde{v}$ refers to this new frame.  From now on we will drop the tilde and only use this new frame.  We now have $\mathbf{u}_1=(\gamma_l, -M, -M \tan  \alpha, 1, 0, \gamma_l)^t$ and $\mathbf{u}_5=(\eta \gamma_l, -M, -M tan  \alpha, 1, 0, \gamma_r)^t$.  The Rankine-Hugoniot relations now immediately give a unique solution for $\mathbf{u}_2$, namely 

\begin{equation}
 \mathbf{u_2}= \left( \frac{(\gamma_l^2 +\gamma_l) M^2}{(\gamma_l-1) M^2 + 2}, - \frac{{(\gamma_l-1) M^2 + 2}}{(\gamma_l +1)M}, -M tan \alpha, \frac{2 \gamma_l M^2 - \gamma_l +1}{\gamma_l+1},0,\gamma_l \right) ^t.
\label{u2}
\end{equation}
In MHD, we nondimensionalise by definining $B=1$ and $\rho = \frac{\gamma_l \beta}{2}$, in region $1$. Again all velocity components are scaled with respect to the sound speed in this region.  We now have that $\mathbf{u}_1 = \left( \frac{\gamma_l \beta}{2}, -M, -M tan \alpha, \frac{\beta + 1}{2}, 1, \gamma_l \right)^t$ and from the definition of $\eta$, $\mathbf{u}_5 = \left( \frac{\eta \gamma_l \beta}{2}, -M, -M tan \alpha, \frac{\beta + 1}{2}, 1 , \gamma_r \right)^t$.  The Rankine-Hugoniot relations now give the following non-trivial solutions for $\mathbf{u}_2$:

\begin{equation}
\mathbf{u}_2 = \left( 
\frac{-\gamma_l \beta M}{2 \omega},\omega,-M tan \alpha ,p_2 + \frac{M^2}{2 \omega^2},\frac{-M}{\omega},\gamma_l \right)^t,
\label{u2mhd}
\end{equation}
where
\begin{equation}
p_2= \frac{A \omega + B}{C \omega + D},
\end{equation}
is the thermal pressure in the post shock region.  We introduced the coefficients
\begin{eqnarray}
A &=&\gamma_l \left( \beta^2 (4\gamma_l^2 M^4 - 2 \gamma_l M^2 - \gamma_l - 1) +\beta \left( (\gamma_l^2 + 4 \gamma_l -5) M^2 -2 \right) -\gamma_l + 2\right),\\
B&=&(\gamma_l - 1)M \left( \beta (M^2 (\gamma_l^2 + 7 \gamma_l) - 2\gamma_l + 4) -2 \gamma_l +4  \right), \\
C&=&2 \gamma_l (\gamma_l +1) \left( \beta ((\gamma_l - 1) M^2 + 2) + 2 \right),\\
D&=& 4(\gamma_l + 1)( \gamma_l -2) M.
\end{eqnarray}
The quantity
\begin{equation}
 \omega = \omega_{\pm} \equiv - \frac{\gamma_l(\gamma_l - 1) \beta M^2 + 2 \gamma_l (\beta + 1) \pm \sqrt{ W}}{2 \gamma_l (\gamma_l + 1) \beta M},
\label{eq:omega}
\end{equation}
is the normal post-shock velocity relative to the shock, with
\begin{eqnarray}
W & = & \beta^2  M^2 (\gamma_l^3 -\gamma_l^2) \left( M^2 (\gamma_l - 1) + 4 \right)  + \beta \gamma_l (4 M^2 (4 + \gamma_l - \gamma_l^2) + 8 \gamma_l) + 4 \gamma_l^2.
\label{eq:W}
\end{eqnarray}
Note that $\omega$ must satisfy $-M < \omega < 0$ to represent a genuine right moving shock.  We choose the solution where $\omega = \omega_{+}$, since the alternative, $\omega = \omega_-$ is a degenerate solution in the sense that the hydrodynamical limit  $ \mathop {\lim }\limits_{\beta \to +\infty} \omega_- = 0$, which does not represent a rightmoving shock.

\subsection{Relations across a contact discontinuity and an expansion fan}

Rewriting equation (\ref{eq:statmhd}) in \textit{quasilinear form} leads to

\begin{equation}
\mathbf{u}_x + \left(\mathbf{F_u}^{-1}\cdot \mathbf{G_u} \right) \mathbf{u}_y = \mathbf{0}.
\label{eq:quastat}
\end{equation}
In the frame moving with the triple point, we are searching for selfsimilar solutions and we can introduce $\xi = \frac{y}{x} = \tan \phi$, so that $\mathbf{u} = \mathbf{u}(\xi)$.  Assuming that $\xi \mapsto \mathbf{u}(\xi)$ is differentiable, manipulating (\ref{eq:quastat}) leads to $\mathbf{A} \mathbf{u}_\xi = \xi \mathbf{u}_\xi$.  So the eigenvalues $\lambda_i$ of $\mathbf{A}$ represent $\tan \phi$, where $\phi$ is the angle between the refracted signals and the negative $x$-axis.  The matrix $\mathbf{A}$ is given by

\begin{equation}
  \mathbf{A} \equiv \mathbf{F}_u^{-1}\mathbf{G}_u=
\left(
     \begin{array}{cccccc}
\frac{v_y}{v_x}& \frac{\rho v_y}{v_x^2-c^2}& -\frac{\rho v_x}{v_x^2-c^2}& \frac{v_y}{v_x}\frac{1}{v_x^2-c^2}&0&0\\
0&  \frac{v_x v_y}{v_x^2-c^2}& -\frac{c^2}{v_x^2-c^2}& -\frac{v_y}{\rho}\frac{1}{v_x^2-c^2}&0&0\\
0&0& \frac{v_y}{v_x}&\frac{1}{\rho v_x}&0&0\\
0 & -\frac{\rho c^2 v_y}{v_x^2-c^2}& \frac{\rho c^2 v_x}{v_x^2-c^2}& \frac{v_x v_y}{v_x^2-c^2}&0&0\\
0& -\frac{B v_y}{v_x^2-c^2}& -\frac{B v_x}{v_x^2-c^2}& \frac{v_y}{v_x}\frac{B}{\rho}\frac{1}{v_x^2-c^2}&\frac{v_y}{v_x}&0\\
0&0&0&0&0&\frac{v_y}{v_x}\\
     \end{array}
\right).
\label{eq:A}
\end{equation}
and its eigenvalues are 

\begin{equation}
 \mathbf{\lambda}_{1,2,3,4,5,6} = \{ \frac{v_x v_y + c \sqrt{v^2 - c^2}}{v_x^2 - c^2} ,\frac{v_y}{v_x}, \frac{v_y}{v_x},\frac{v_y}{v_x},\frac{v_y}{v_x},\frac{v_x v_y - c \sqrt{v^2 - c^2}}{v_x^2 - c^2} \},
\label{eq:eigenvalues}
\end{equation}
where the magnetosonic speed $c \equiv \sqrt{v_s^2 + v_a^2}$ and the sound speed $v_s = \sqrt{\frac{\gamma p}{\rho}}$ and the Alfv\'en speed $v_a = \sqrt\frac{B^2}{\rho}$.  Since $\mathbf{A}$ has $3$ different eigenvalues, $3$ different signals will arise.  When $\mathbf{u}_{\xi}$ exists and $\mathbf{u}_{\xi} \neq \mathbf{0}$, i.e. inside of expansion fans, $\mathbf{u}_\xi$ is proportional to a right eigenvector $\mathbf{r}_i$ of $\mathbf{A}$.  Derivation of $\xi = \lambda_i$ with respect to $\xi$ gives $(\nabla_{\mathbf{u}} \lambda_i)\cdot \mathbf{u}_\lambda =1$ and thus we find the proportionality constant, giving 

\begin{equation}
 \mathbf{u}_\xi = \frac{\mathbf{r_i}}{\nabla_{\mathbf{u}} \lambda_i \cdot \mathbf{r_i}}. 
\label{uxi}
\end{equation}

While this result assumed continuous functions, we can also mention relations that hold even across discontinuities like the $CD$.  Denoting the ratio $\frac{d\mathbf{u}_i}{\mathbf{r_i}} = \kappa$, it follows that $\left[ \mathbf{l}_i \cdot d\mathbf{u}\right]_{dx=\lambda_j dy} = ( \mathbf{l}_i \cdot \mathbf{r}_j)\kappa = \kappa \delta_{i,j}$, where $\mathbf{l_i}$ and $\mathbf{r_i}$ are respectively left and right eigenvectors corresponding to $\lambda_i$.   Therefore, if $i \neq j$,
\begin{equation}
 \left[ \mathbf{l}_i \cdot d\mathbf{u}\right]_{dx=\lambda_j dy} = 0.
\label{eq:li}
\end{equation}
From these general considerations the following relations hold across the contact or shear wave where the ratio $\frac{dy}{dx}=\frac{v_y}{v_x}$ :
\begin{equation}
 \left\lbrace \begin{array}{c}
 v_y d v_x - v_x d v_y + \frac{ c \sqrt{v^2-c^2}}{\rho v_s^2} dp_{tot} = 0,\\
 v_y d v_x - v_x d v_y - \frac{ c \sqrt{v^2-c^2}}{\rho v_s^2} dp_{tot} = 0.\\
\end{array} \right. 
\label{eq:RI2}
\end{equation}
Since $v \neq c$, otherwise all signals would coincide, it follows immediately that the total pressure $p_{tot}$ and the direction of the streamlines $\frac{v_y}{v_x}$ remain constant across the shocked contact discontinuity.

These relations across the CD allow to solve the full problem using an iterative procedure.  Inspired by the exact Riemann solver described in \cite{TO99}, we first guess the total pressure $p^*$ across the $CD$.  $R$ is a shock when $p^*$ is larger than the post-shock total pressure and  $T$ is a shock, only if $p^*$ is larger than the pre-shock total pressure.  Note that the jump in tangential velocity aross the CD is a function of $p^*$ and it must vanish.  A simple Newton-Raphson iteration on this function $[[\frac{v_y}{v_x}]](p^*)$, finds the correct $p^*$.  We explain further in section $3.5$ how we find the functional expression and iterate to eventually quantify $\phi_{R}$, $\phi_{T}$, $\phi_{CD}$ and the full solution $\mathbf{u}(x,y,t)$.  From now on $p^*$ represents the constant total pressure across the CD.

Similarly, from the general considerations above, equation (\ref{eq:li}) gives that along $\frac{dy}{dx}=\frac{v_x v_y \pm c \sqrt{v^2 - c^2}}{v_x^2 - c^2}$ the following relations connect two states across expansion fans:
\begin{equation}
 \left\lbrace \begin{array}{c}
 d \rho -\frac{1}{c^2} dp_{tot} = 0,\\
 v_x d v_x +v_y d v_y + \frac{c^2}{\rho v_s^2} dp_{tot} = 0,\\
-\rho dp_{tot} + p_{tot} \rho d\gamma +p_{tot} \gamma d\rho = 0,\\
-B d p_{tot} + \left(\gamma p + B^2 \right) dB=0,\\
 v_y d v_x - v_x d v_y \pm \frac{ c \sqrt{v^2-c^2}}{\rho v_s^2} dp_{tot} = 0.\\
\end{array} \right.
\label{eq:RI}
\end{equation}
These can be written in a form which we exploit to numerically integrate the solution through expansion fans, namely
\begin{equation}
\left\lbrace \begin{array}{c}
\rho_i = \rho_e + \int_{p_{tot,e}}^{p^*} \frac{1}{c^2} dp_{tot},\\
v_{x,i} = v_{x,e}+\int_{p_{tot,e}}^{p^*} \frac{\pm v_y \sqrt{v^2-c^2}-v_x c}{\rho v^2 c}dp_{tot},\\
v_{y,i} = v_{y,e}+\int_{p_{tot,e}}^{p^*} \frac{\mp v_x \sqrt{v^2-c^2}-v_y c}{\rho v^2 c}dp_{tot},\\
B_i=B_e+\int_{p_{tot,e}}^{p^*} \frac{B}{\rho c^2} d p_{tot},\\
p_i=p_e+\int_{p_{tot,e}}^{p^*} \frac{v_s^2}{ c^2} d p_{tot},\\
\gamma_i = \gamma_e.
\end{array} \right. 
\label{eq:expansionintegration}
\end{equation}
The indices $i$ and $e$ stand respectively for \textit{internal} and \textit{external}, the states at both sides of the expansion fans.  The upper signs hold for reflected expansion fans (i.e. of type R), while the lower sign holds for transmitted expansion fans (i.e. of type T).

\subsection{Relations across a shock}

Since the system is nonlinear and allows for large-amplitude shock waves, the analysis given thus far is not sufficient.  We must include the possibility of one or both of the R and T signals to be solutions of the stationary Rankine-Hugoniot conditions (equation (\ref{eq:statrh})).  The solution is given by

\begin{equation}
\left\lbrace
\begin{array}{c}
\rho_{i}=\frac{\frac{\gamma-1}{\gamma+1}+\frac{p^*}{p_{tot,e}}}{\frac{\gamma-1}{\gamma+1}\frac{p^*}{p_{tot,e}}+1}\rho_e,\\
v_{x,i}=v_{x,e}-\frac{\xi_\mp(p^*-p_{tot,e})}{\rho_e (v_{x,e} \xi_\mp - v_{y,e})},\\
v_{y,i}=v_{y,e}+\frac{p^*-p_{tot,e}}{\rho_e (v_{x,e} \xi_\mp - v_{y,e})},\\
B_i= \frac{\frac{\gamma-1}{\gamma+1}+\frac{p^*}{p_{tot,e}}}{\frac{\gamma-1}{\gamma+1}\frac{p^*}{p_{tot,e}}+1} B_e,\\
\gamma_i = \gamma_e,\\
p_i=p^*-\frac{B_i^2}{2},\\
\phi_{R/T}=atan(\xi_{+/-}),\\
\end{array}
\right.
\label{{eq:u3}}
\end{equation}
where 
\begin{equation}
\xi_{\pm} =\frac{v_{e,x} v_{e,y} \pm \hat{c}_{e} \sqrt{v_e^2 - \hat{c}_e^2}}{v_{e,x}^2 - \hat{c}_e^2},
\label{eq:xipm}
\end{equation}
and
\begin{equation}
\hat{c}_e^2=\frac{(\gamma-1)p_{tot,e} + (\gamma + 1)p^*}{2 \rho_e}.
\label{eq:chat}
\end{equation}
Again the indices $i$ and $e$ stand respectively for \textit{internal} and \textit{external}, the states at both sides of the shocks.  The upper signs holds for reflected shocks, while the lower sign holds for transmitted shocks.

\subsection{Shock refraction as a Riemann problem}

We are now ready to formulate our iterative solution strategy.  Since there exist $2$ invariants across the CD, it follows that we can do an iteration, if we are able to express one invariant in function of the other.  As mentioned earlier, we choose to iterate on $p^*=p_{tot,3}=p_{tot,4}$.  This is the only state variab;e in the solution, and it controls both R and T.  We will write $\phi_{R}=\phi_{R}(\mathbf{u}_{2},p*)$ and $\phi_{T}=\phi_{T}(\mathbf{u}_{5},p^*)$, $\mathbf{u}_{3}=\mathbf{u}_{3}(\mathbf{u}_{2},p^*)$ and $\mathbf{u}_{4}=\mathbf{u}_{4}(\mathbf{u}_{5},p^*)$.  The other invariant should match too, i.e. $\frac{v_{x,3}}{v_{y,3}} - \frac{v_{x,4}}{v_{y,4}} = 0$.  Since $\mathbf{u}_2$ and $\mathbf{u}_5$ only depend on the input parameters, this last expression is a function of $p^*$.  Iteration on $p^*$ gives $p^*$ and $\phi_{R}=\phi_{R}(p^*)$, $\phi_{T}=\phi_{T}(p^*)$, $\mathbf{u}_{3}=\mathbf{u}_{3}(p^*)$ and $\mathbf{u}_{4}=\mathbf{u}_{4}(p^*)$  give $\phi_{CD}= atan{\frac{v_{y,3}}{v_{x,3}}} = atan{\frac{v_{y,4}}{v_{x,4}}}$, which solves the problem.

\subsection{Solution inside of an expansion fan}
The only ingredient not yet fully specified by our description above is how to determine the variation through possible expansion fans.  This can be done once the solution for $p^*$ is iteratively found, by integrating equations (\ref{eq:expansionintegration}) till the appropriate value of $p_{tot}$.  Notice that the location of the tail of the expansion fan is found by $tan(\phi_{tail})=\frac{v_{y,i} v_{x,i} \pm c_i \sqrt{v_i^2 - c_i^2}}{v_{x,i}^2 -c_i^2}$ and the position of $\phi_{head}$ is uniquely determined by  $tan(\phi_{head})=\frac{v_{y,e} v_{x,e} \pm c_e \sqrt{v_e^2 - c_e^2}}{v_{x,e}^2 -c_e^2}$.  Inside an expansion fan we know $\mathbf{u}(p_{tot})$, so now we need to find $p_{tot}(\phi)$, in order to find a solution for $\mathbf{u}(\phi)$.  We decompose vectors locally in the normal and tangential directions, which are  respectively referred to with the indices $n$ and $t$.  We denote taking derivatives with respect to $\phi$ as $'$.
Inside of the expansion fans we have some invariants given by equations (\ref{eq:RI}).  The fourth of these immediately leads to $\frac{p}{B^{\gamma}}$ as an invariant.  Eliminating $p_{tot}$ from  $d \rho -\frac{1}{c^2} dp_{tot} = 0$ and $-B d p_{tot} + (\gamma p +B^2)dB=0$ yields the invariant $\frac{\rho}{B}$, and combining these 2 invariants tells us that the entropy $S \equiv \frac{p}{\rho^\gamma}$ is invariant.
The stationary MHD equations (\ref{eq:statmhd}) can then be written in a $4 \times 4$-system for $v'_n, v'_t, p'_{tot}$ and $\rho'$ as:
\begin{equation}
\left\lbrace \begin{array}{c}
v'_n+v_t+v_n \frac{\rho'}{\rho}=0,\\
v_n v_t+ v_n v'_n +\frac{p'_{tot}}{\rho}=0,\\
v_n^2-v_n v'_t=0,\\
c^2 \rho' - p'_{tot}=0,\\
\end{array} \right.
\label{eq:infan}
\end{equation}
where we dropped $B'$ from the system, since it is proportional to $\rho'$.  Note that $\gamma'$ vanishes.  The system leads to the dispersion relation
\begin{equation}
v_n^4-c^2 v_n^2 = 0,
\label{eq:dispersion}
\end{equation}
which in differential form becomes:
\begin{equation}
4 \rho v_n^3 v'_n +v_n^4 \rho' -\gamma v_n^2 p'_{tot} - 2 \gamma p_{tot} v_n v'_n -(2-\gamma)B v_n^2 B' -(2-\gamma)B^2 v_n v'_n = 0.
\label{eq:dispder}
\end{equation}
Elimination of $v'_n$, $\rho'$ and $B'$ gives
\begin{equation}
\frac{dp_{tot}}{d \phi} =  2 \frac{v_t}{v_n} \frac{c^2  - 2 v_n^2}{3v_n^2+(\gamma-2)c^2} \rho c^2.
\end{equation}
This expression allows us to then complete the exact solution as a function of $\phi$.

\section{Implementation and numerical details}

\subsection{Details on the Newton-Raphson iteration}

We can generally note that $p_{tot,pre} < p_{tot,post}$.  This implies that the refraction has $3$ possible wave configurations:  $2$ shocks, a reflected rarefaction fan and a transmitted shock, or $2$ expansion fans.  Before starting the iteration on $[[\frac{v_y}{v_x}]](p^*)$, we determine the governing wave configuration.  If $[[\frac{v_y}{v_x}]](\epsilon)$ and $[[\frac{v_y}{v_x}]](p_{tot,5} -\epsilon)$ differ in sign, the solution has two rarefaction waves.  If $[[\frac{v_y}{v_x}]](p_{tot,5} + \epsilon)$ and $[[\frac{v_y}{v_x}]](p_{tot,2}-\epsilon)$ differ in sign, the solution has a transmitted shock and a reflected rarefaction wave.  In the other case, the solution contains two shocks in its configuration.
If $R$ is an expansion fan, we take the guess \begin{equation}p_0^*=\frac{min\{\frac{2 \rho_e v_{x,e}^2 - (\gamma_e - 1) p_{tot,e}}{\gamma+1} |e\in\{2,5\}\} + p_{tot,5}}{2} \end{equation} as a starting value of the iteration.  This guess is the mean of the critical value $p_{tot,crit}$ , which satisfies 
\begin{equation}
v_{e,x}^2 -\hat{c}^2(p_{tot,crit})=0,
\end{equation}
 and $p_5$, which is the minimal value for a transmitted shock.  As we explain in section $5.3$,  $v_{2,x}^2 -\hat{c}^2(p_{tot,crit})=0$ is equivalent to $v_{5}^2 -\hat{c}^2 = 0$ and $v_{5,x}^2 -\hat{c}^2(p_{tot,crit})=0$ is equivalent to $v_{2}^2 -\hat{c}^2 = 0$, and is thus a maximal value for the existence of a regular solution.  If $R$ is a shock, we take $(1+\hat{\epsilon})p_{post}$ as a starting value for the iteration, where $\hat{\epsilon}$ is $10^{-6}$.
We use a Newton-Raphson interation: $p^*_{i+1}=p^*_i - \frac{f(p^*_i)}{f'(p^*_i)}$, where $f'(p^*)$ is approximated numerically by $\frac{f(p^*_i + \delta) - f(p^*_i)}{\delta}$, where $\delta =10^{-8}$.  The iteration stops when $\frac{p^*_{i+1}-p^*_i}{p^*_i} < \epsilon$, where $\epsilon = 10^{-8}$.

\subsection{Details on AMRVAC}
\begin{figure}
\centering
 \includegraphics[width=.99\textwidth]{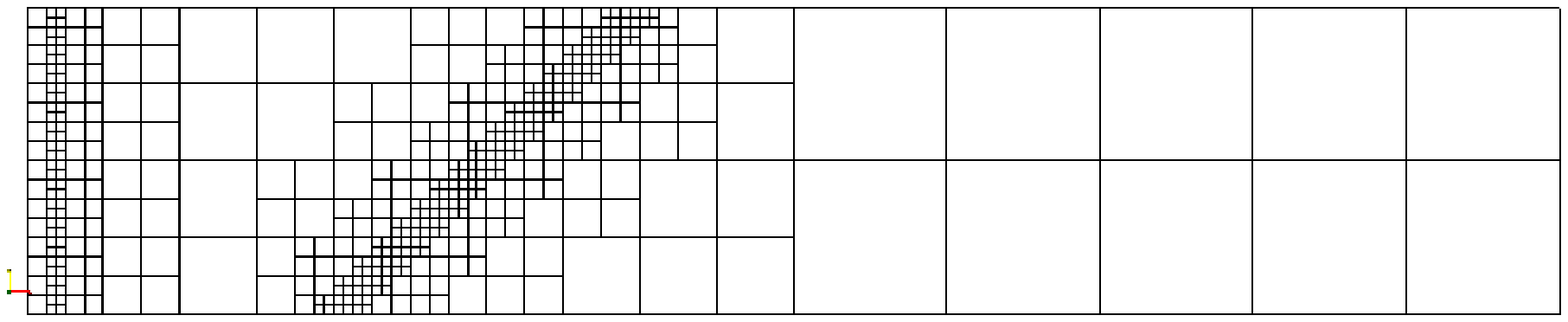}
\caption{The initial AMR grid at $t=0$, for the example in section $5.1$.}
\label{fig:amr}
\end{figure}

AMRVAC (\cite{HO07, KE03}) is an AMR code, solving equations of the general form $\mathbf{u}_t +\nabla \cdot \mathbf{F(u)} = \mathbf{S}(\mathbf{u},\mathbf{x},t)$ in any dimensionality.  The applications cover multi-dimensional HD, MHD, up to special relativistic magnetohydrodynamic computations.  In regions of interests, the AMR code dynamically refines the grid.  The initial grid of our simulation is shown in figure~\ref{fig:amr}.  The refinement strategy is done by quantifying and comparing gradients.  The AMR in AMRVAC is of a block-based nature, where every refined grid has $2^D$ children, and $D$ is the dimensionality of the problem.  Parallelisation is implemented, using MPI.  In all the simulations we use $5$ refinement levels, starting with a resolution of $24 \times 120$ on the domain $[0,1] \times [0,5]$, leading to an effective resolution of $384 \times 1940$.  The shock is initially located at $x=0.1$, while the contact discontinuity is located at $y =  (x - 1)tan \alpha$.  We used the fourth order Runge-Kutta timestepping, together with a TVDLF-scheme (see \cite{TO96,YE90}) with Woodward-limiter on the primitive variables.  The obtained numerical results were compared to and in agreement with simulations using other schemes, such as a Roe scheme and the TVD-Muscl scheme.  The calculations were performed on $4$ processors.

\subsection{Following an interface numerically}

The AMRVAC implementation contains slight differences with the theoretical approach.  Implementing the equations as we introduced them here would lead to excessive numerical diffusion on $\gamma$.  Since $\gamma$ is a discrete variable we know $\gamma(x,y,t)$ exactly, if we are able to follow the contact discontinuity in time.  Suppose thus that initially a surface, seperates 2 regions with different values of $\gamma$. Define a function $\chi: D \times \mathbb{R}^+ \rightarrow \mathbb{R} : (x, y, t) \mapsto \chi(x, y, t)$, where $D$ is the physical domain of $(x,y)$.  Writing $\tilde{\chi}(x,y)=\chi(x,y,0)$, we ask $\tilde{\chi}$ to vanish on the initial contact and to be a smooth function obeying

\begin{itemize}
 \item $\gamma = \gamma_l \Leftrightarrow \tilde{\chi}(x,y) < 0$,
 \item $\gamma = \gamma_r \Leftrightarrow \tilde{\chi}(x,y) > 0$.
\end{itemize}
We take in particular $\pm \tilde{\chi}$ to quantify the shortest distance from the point $(x,y)$ to the initial contact, taking the sign into account.  Now we only have to note that $(\chi \rho)_t = \chi \rho_t + \rho \chi_t = -\chi \nabla \cdot (\rho \mathbf{v}) - (\rho \mathbf{v} \cdot \nabla) \chi= -\nabla \cdot (\chi \rho \mathbf{v})$.  The implemented system is thus (~\ref{eq:statmhd}), but the last equation is replaced by  $\left(\chi \rho v_x\right)_x + \left( \chi \rho v_y \right)_y = 0$.  It is now straightforward to show that we did not introduce any new signal.  In essence, this is the approach presented in \cite{MU92}.
\begin{figure}
\centering
a) \includegraphics[width=.32\textwidth,angle=-90]{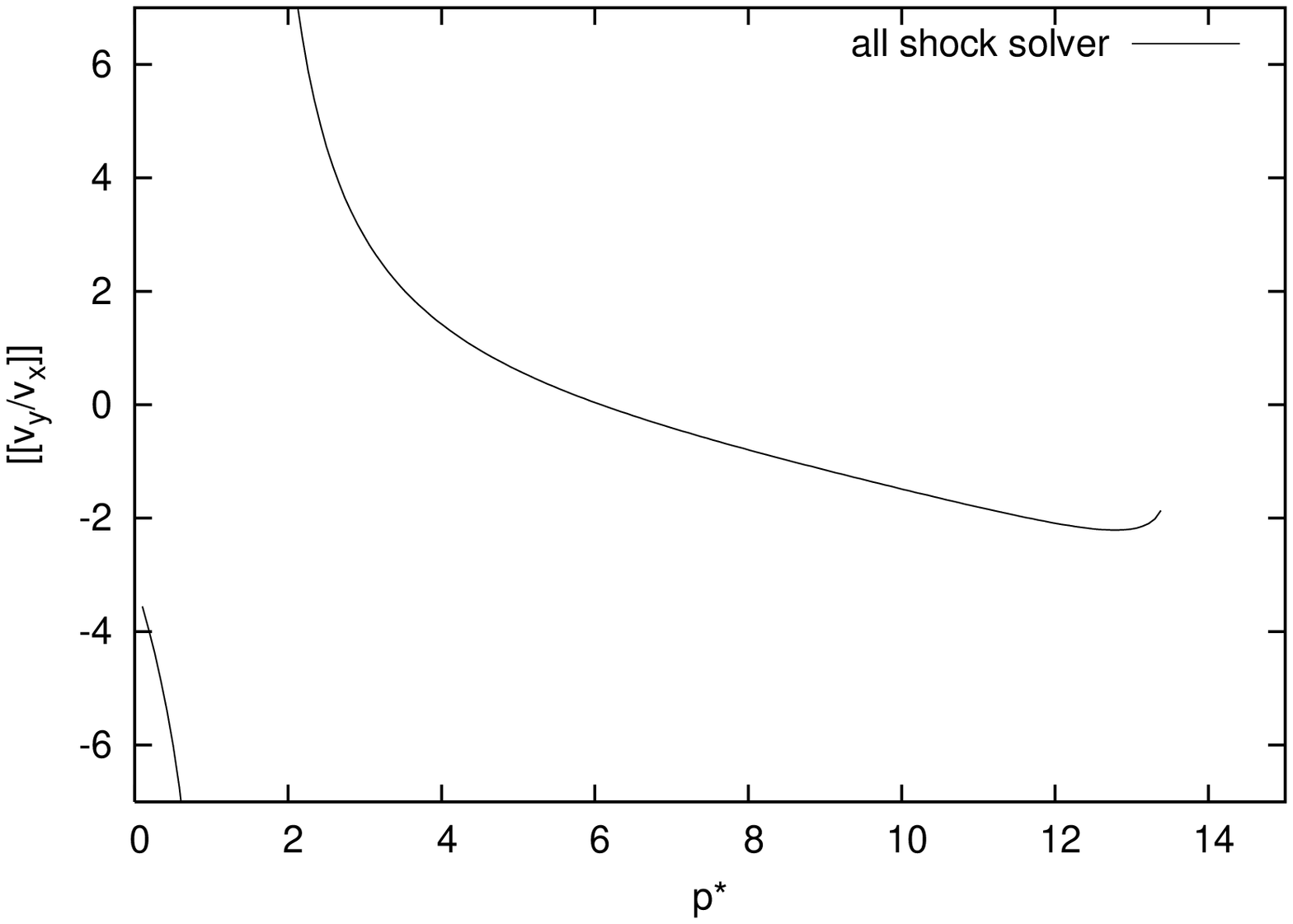}
b) \includegraphics[width=.32\textwidth,angle=-90]{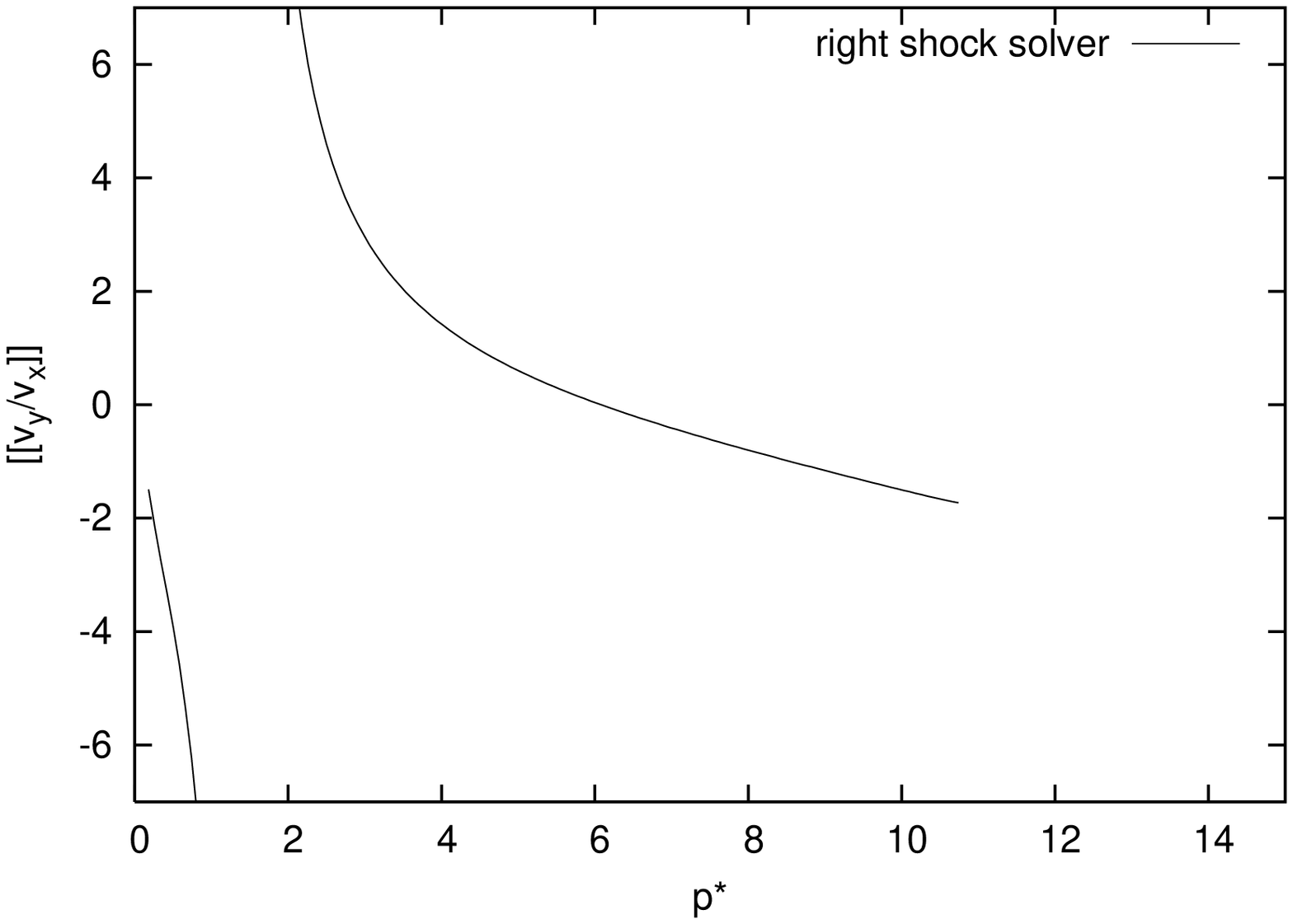}

c) \includegraphics[width=.32\textwidth,angle=-90]{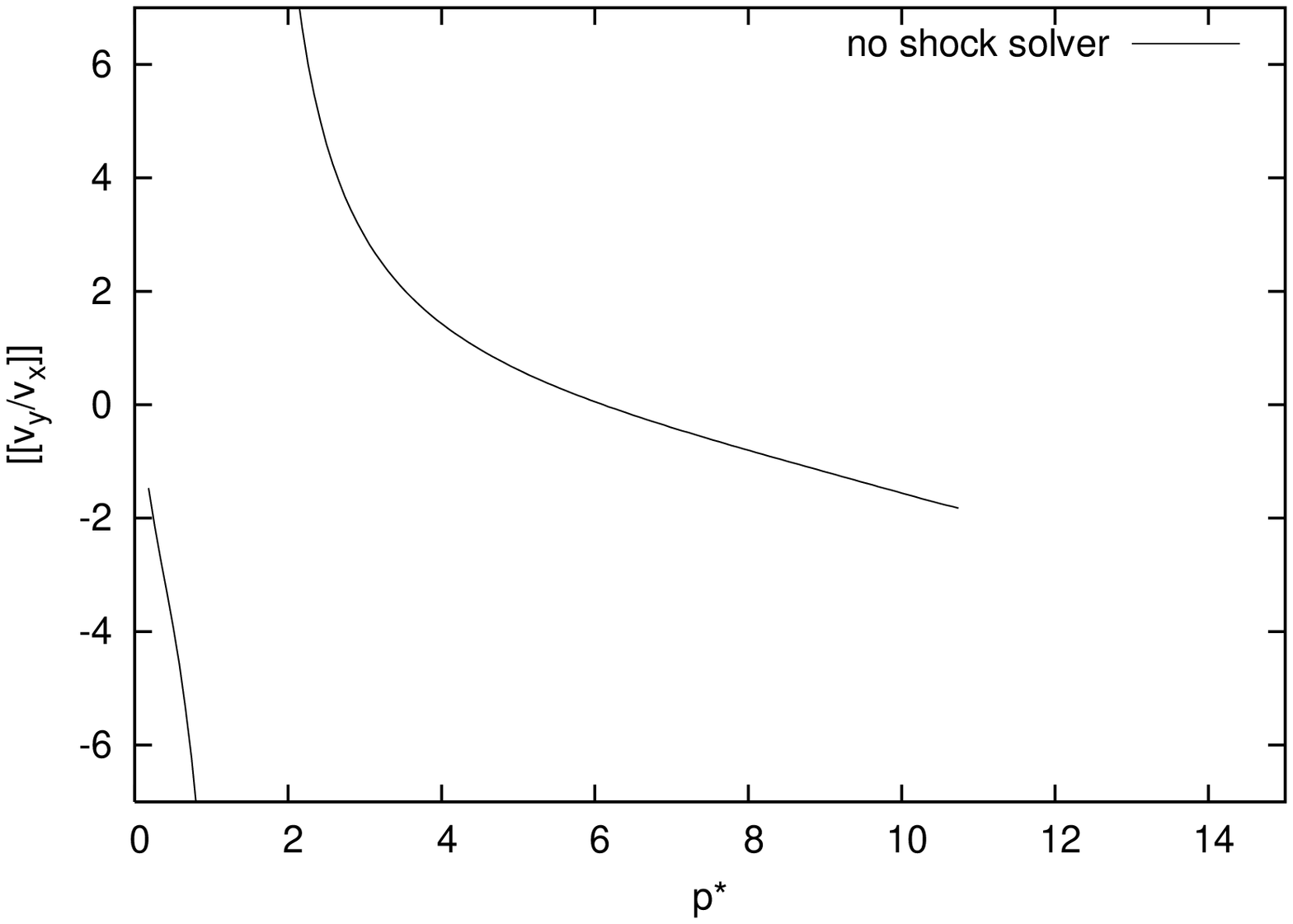}
d) \includegraphics[width=.32\textwidth,angle=-90]{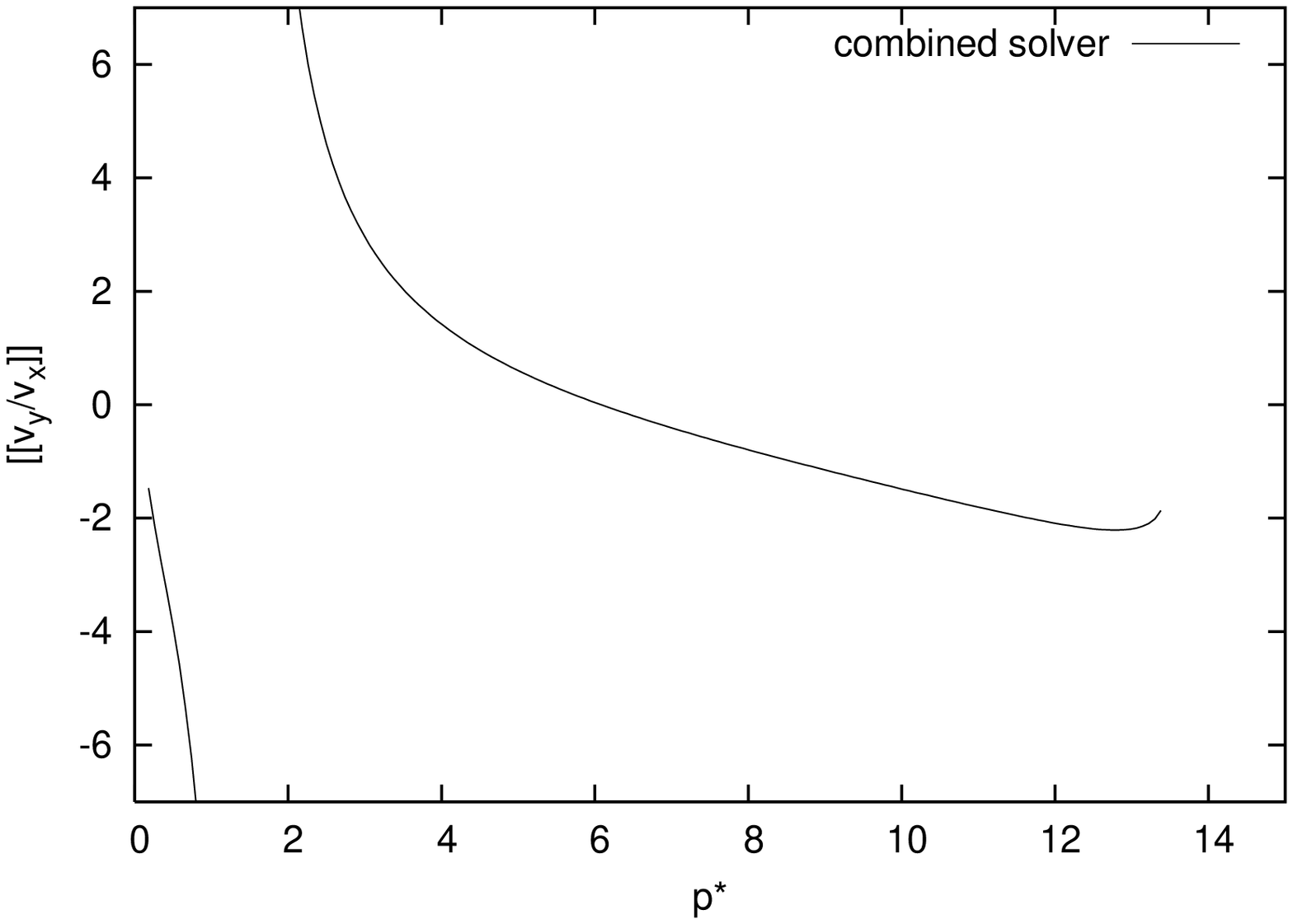}
\caption{$\left[\left[ \frac{v_y}{v_x}\right]\right](p^*)$ for the reference case from \cite{SA03}: a) all shock solver; b) right shock solver; c) no shock solver; d) shock $\Leftrightarrow p^*>p_i$.   The all shock solver is selected.}
\label{fig:solver}
\end{figure}
\section{Results}

\subsection{Fast-Slow example solution}

\begin{figure}
\centering
\includegraphics[width=.32\textwidth,angle=-90]{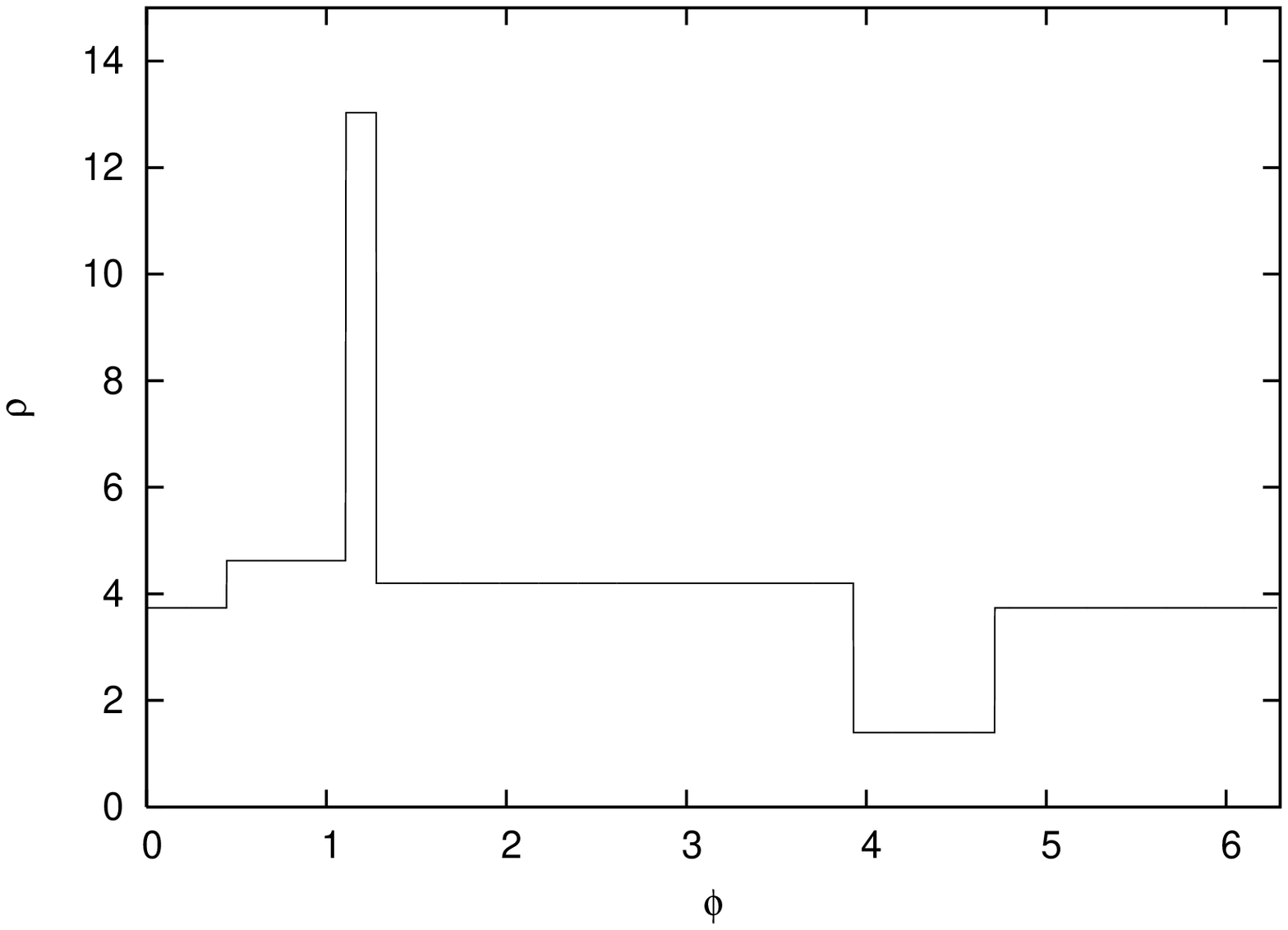}
\includegraphics[width=.32\textwidth,angle=-90]{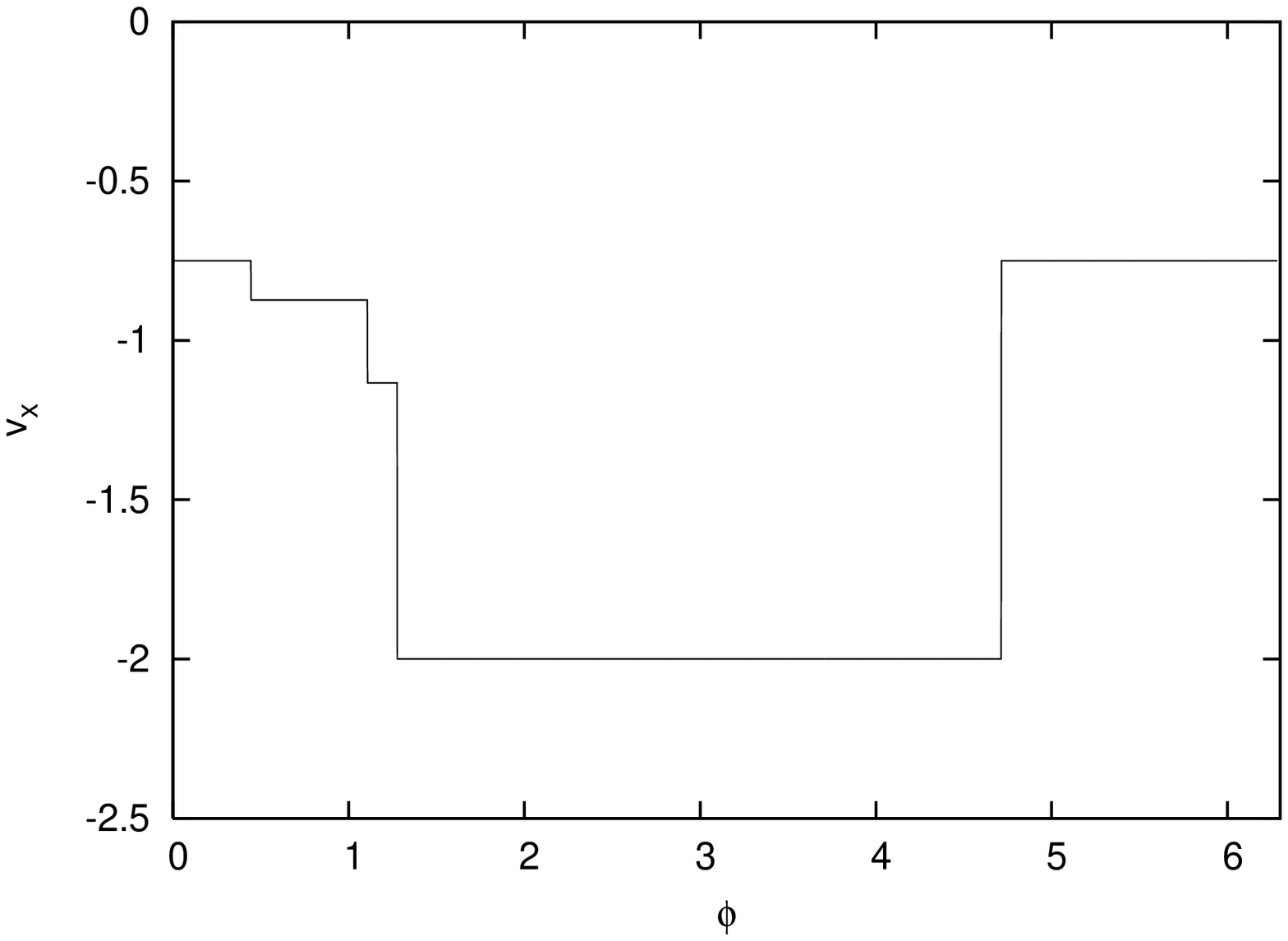}
\includegraphics[width=.32\textwidth,angle=-90]{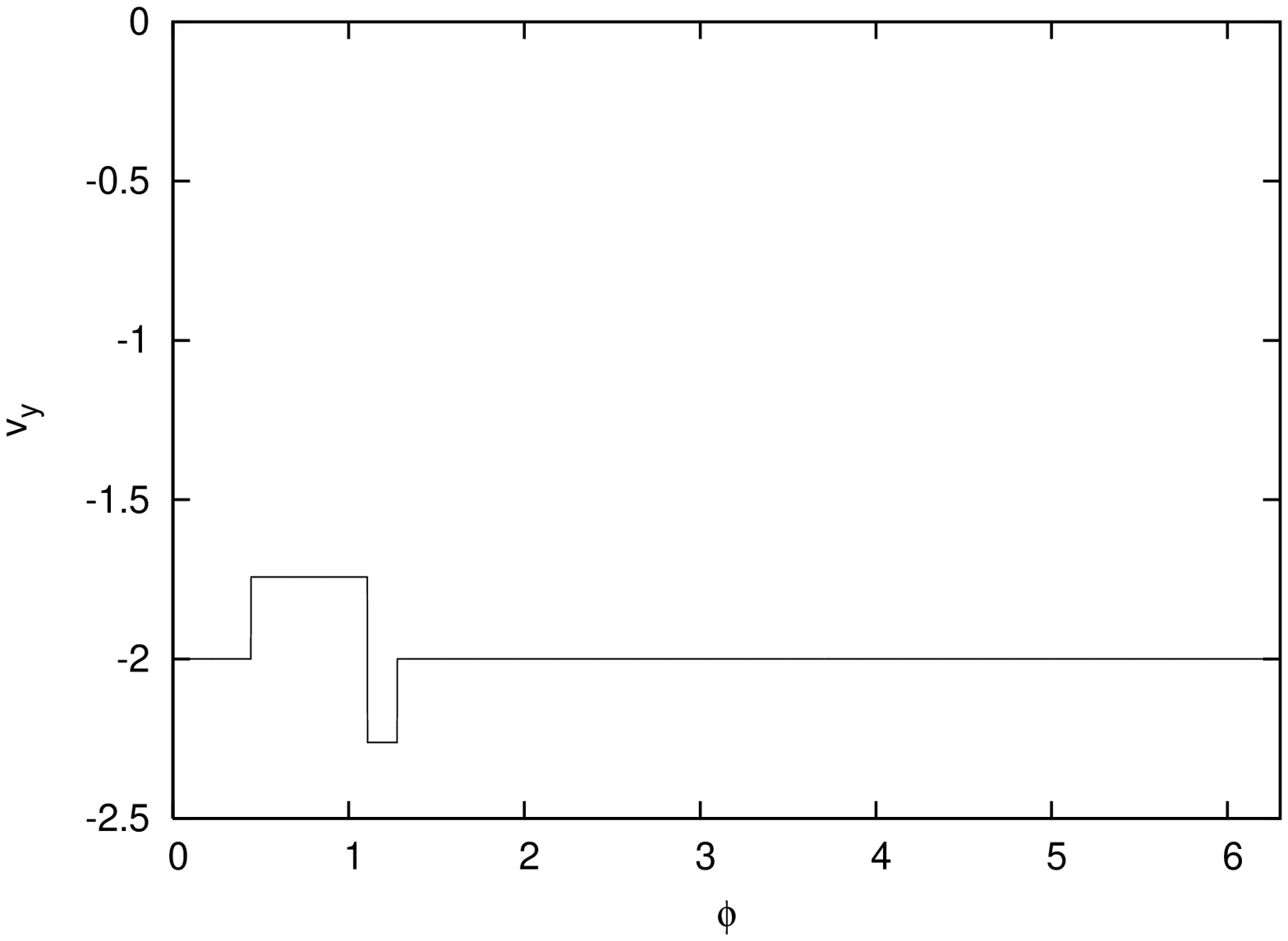}
\includegraphics[width=.32\textwidth,angle=-90]{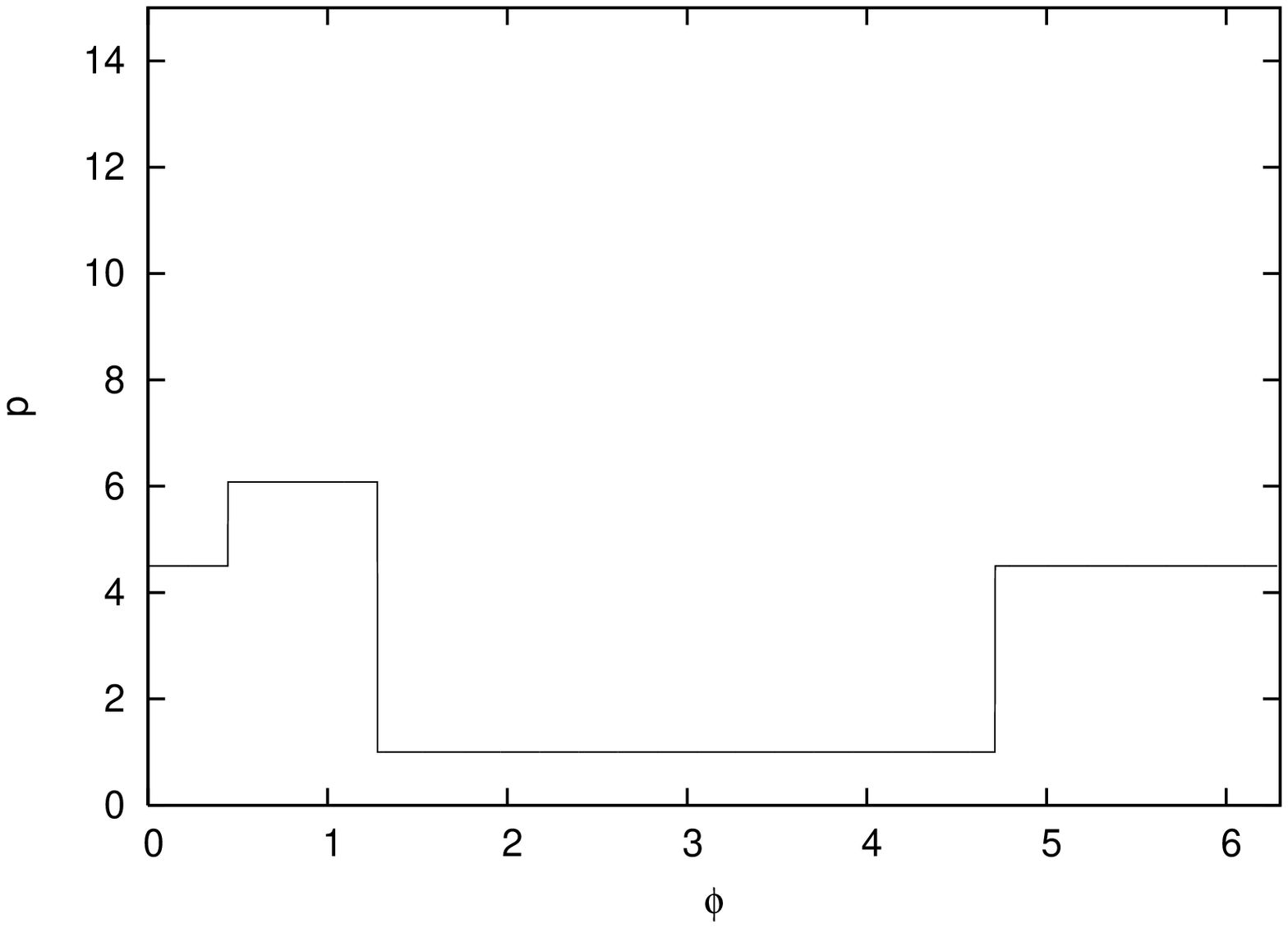}
\includegraphics[width=.32\textwidth,angle=-90]{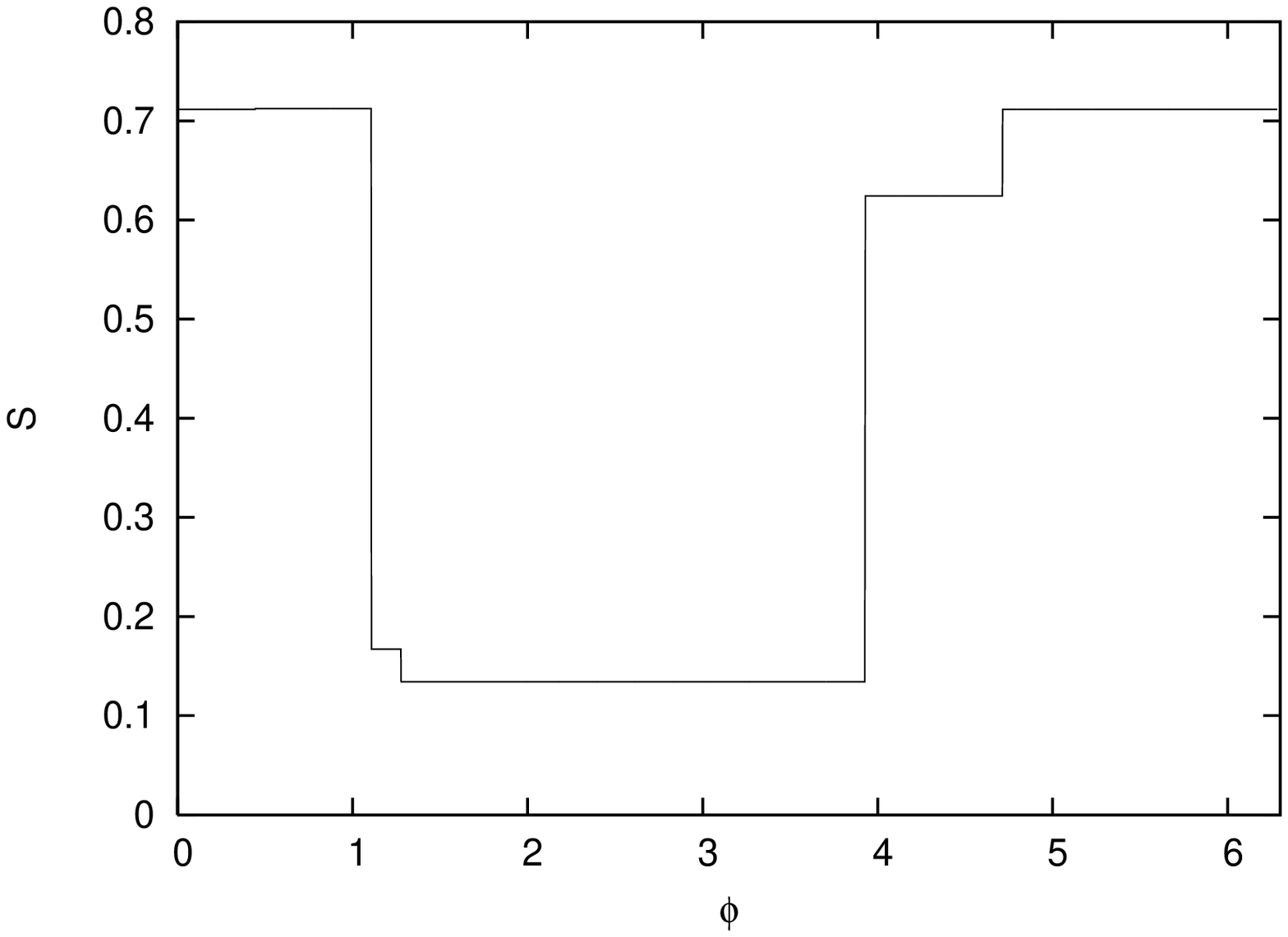}
\includegraphics[width=.32\textwidth,angle=-90]{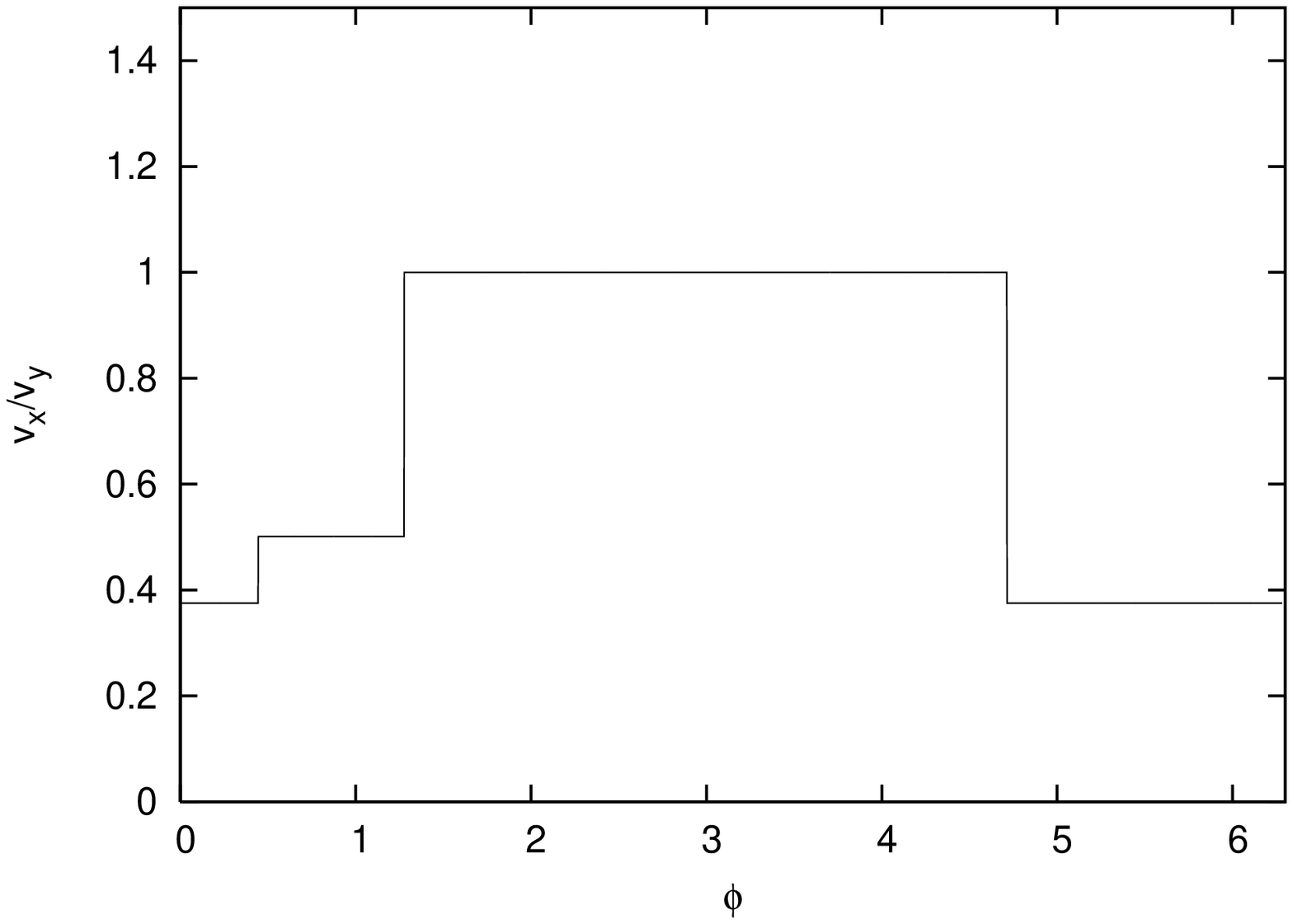}
\caption{Solution to the fast-slowrefraction problem, for the reference case from \cite{SA03}.  Notice that $p$ and $\frac{v_x}{v_y}$ remain constant across the shocked contact.}
\label{fig:sol1}
\end{figure}
As a first hydrodynamical example, we set  $\left( \alpha, \beta^{-1}, \gamma_l, \gamma_r, \eta, M \right) = \left( \frac{\pi}{4}, 0, \frac{7}{5}, \frac{7}{5},3,2 \right)$, as originally presented in \cite{SA03}.  In figure~\ref{fig:solver}, the first 3 plots show $[[\frac{v_y}{v_x}]](p^*)$, when assuming a prescribed wave configuration, for all $3$ possible configurations.   The last plot shows the actual function $[[\frac{v_y}{v_x}]](p^*)$, which consists of piecewise copies from the 3 possible configurations in the previous plots.  The initial guess is $p_0^*=4.111$, the all shock solver is selected, and the iteration converges after $6$ iterations with $p^*=6.078$.  The full solution of the Riemann problem is shown in figure~\ref{fig:sol1}.

\subsection{Slow-Fast example}

\begin{figure}
\centering
\includegraphics[width=.32\textwidth,angle=-90]{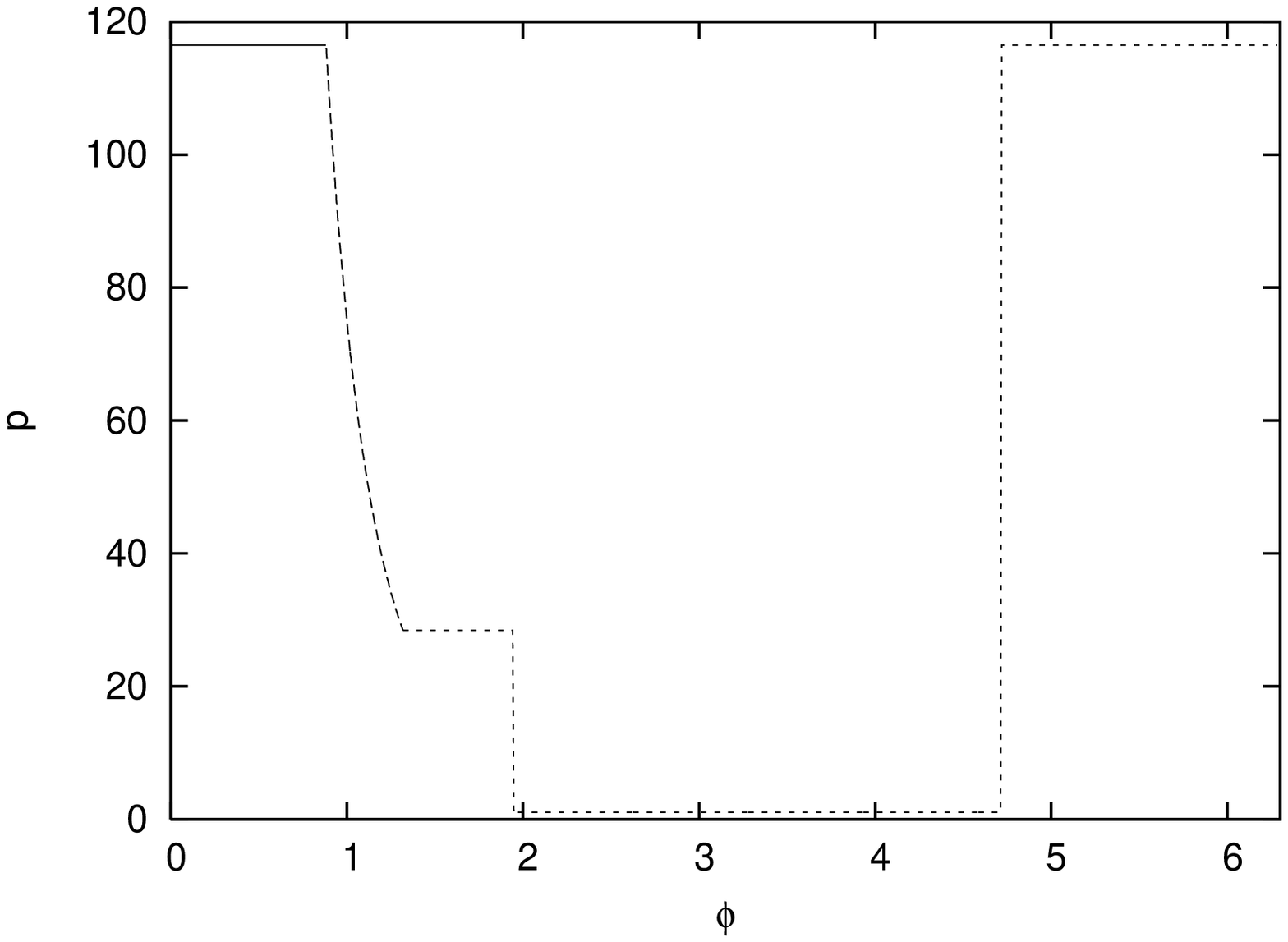}
\includegraphics[width=.32\textwidth,angle=-90]{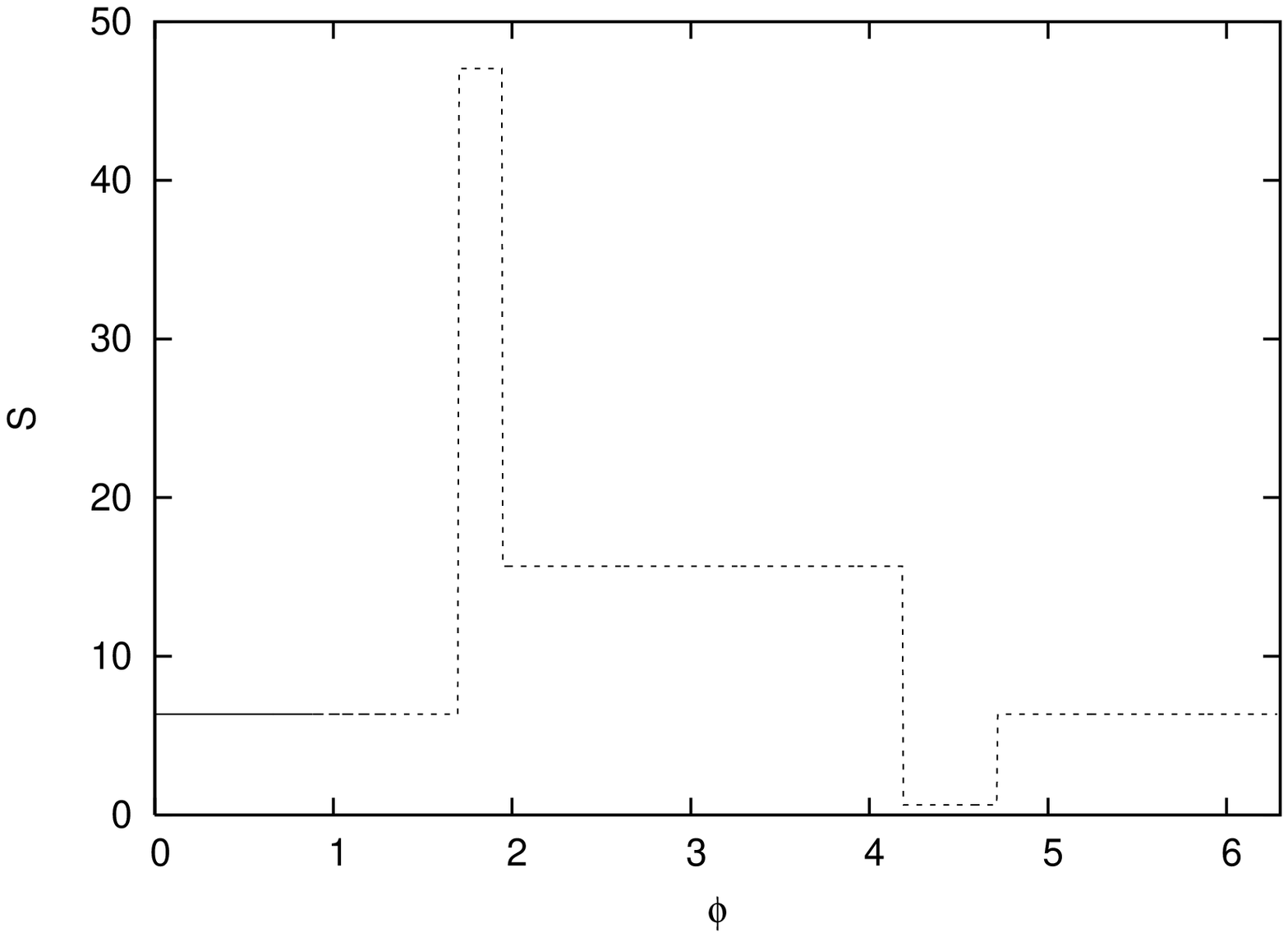}

\caption{Solution to the slow-fast refraction problem from \cite{HO07}.  Notice that $S$ remains constant across $R$.}
\label{fig:sol2}
\end{figure}
In figure~\ref{fig:sol2} we show the full solution of the HD Riemann problem, in which the reflected signal is an expansion fan, connected to the refraction with parameters $\left( \alpha, \beta^{-1}, \gamma_l, \gamma_r, \eta, M \right) = \left( \frac{\pi}{3}, 0, \frac{7}{5}, \frac{7}{5},\frac{1}{10},10 \right)$ from \cite{HO07}.  The refraction is slow-fast, and $R$ is an expansion fan.  Note that $p$ and $\frac{v_y}{v_x}$ remain constant across the CD, and the entropy $S$ is an invariant across $R$.

\subsection{Tracing the critical angle for regular shock refraction}

\begin{figure}
\centering
\includegraphics[width=.32\textwidth,angle=-90]{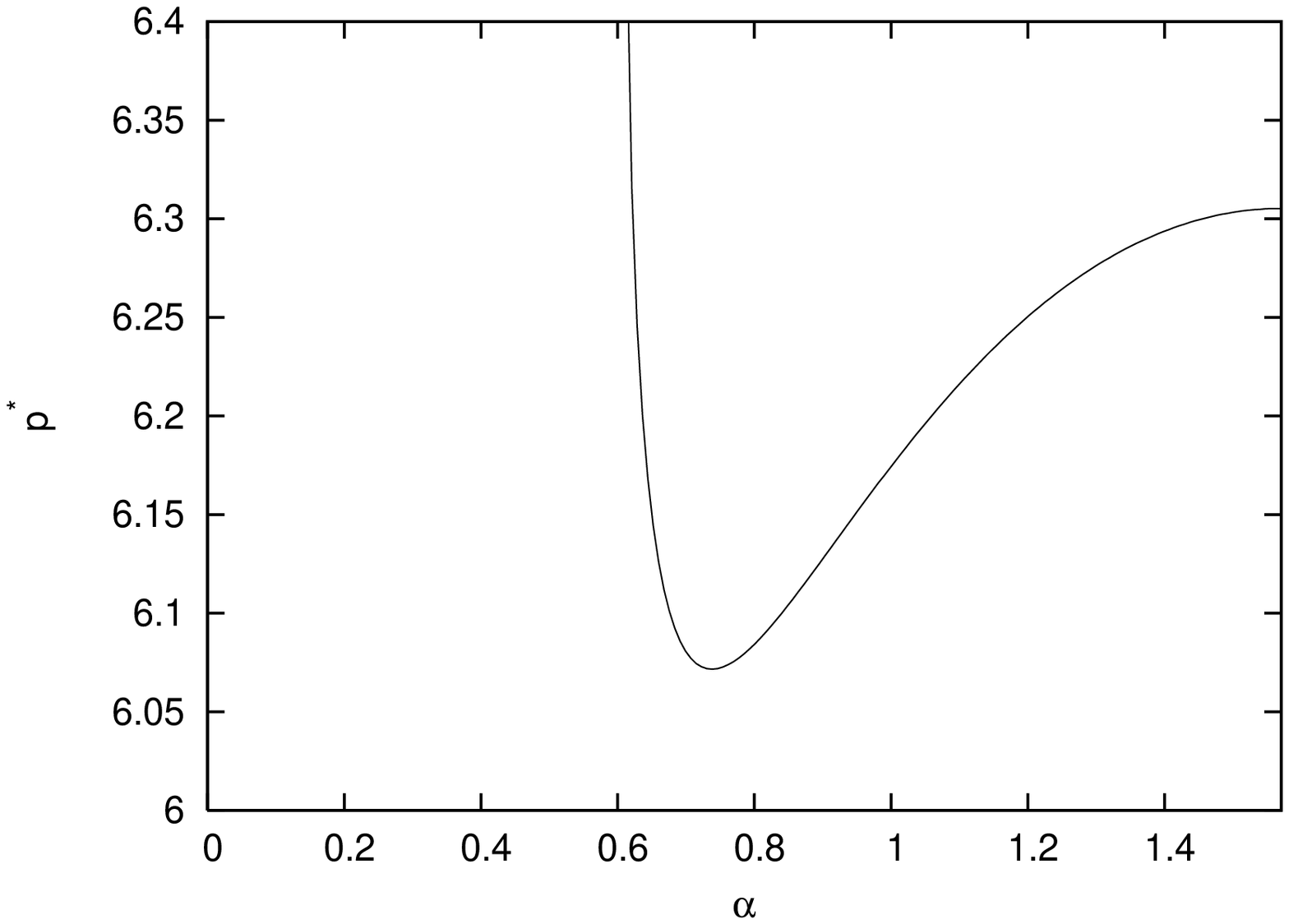}
\includegraphics[width=.32\textwidth,angle=-90]{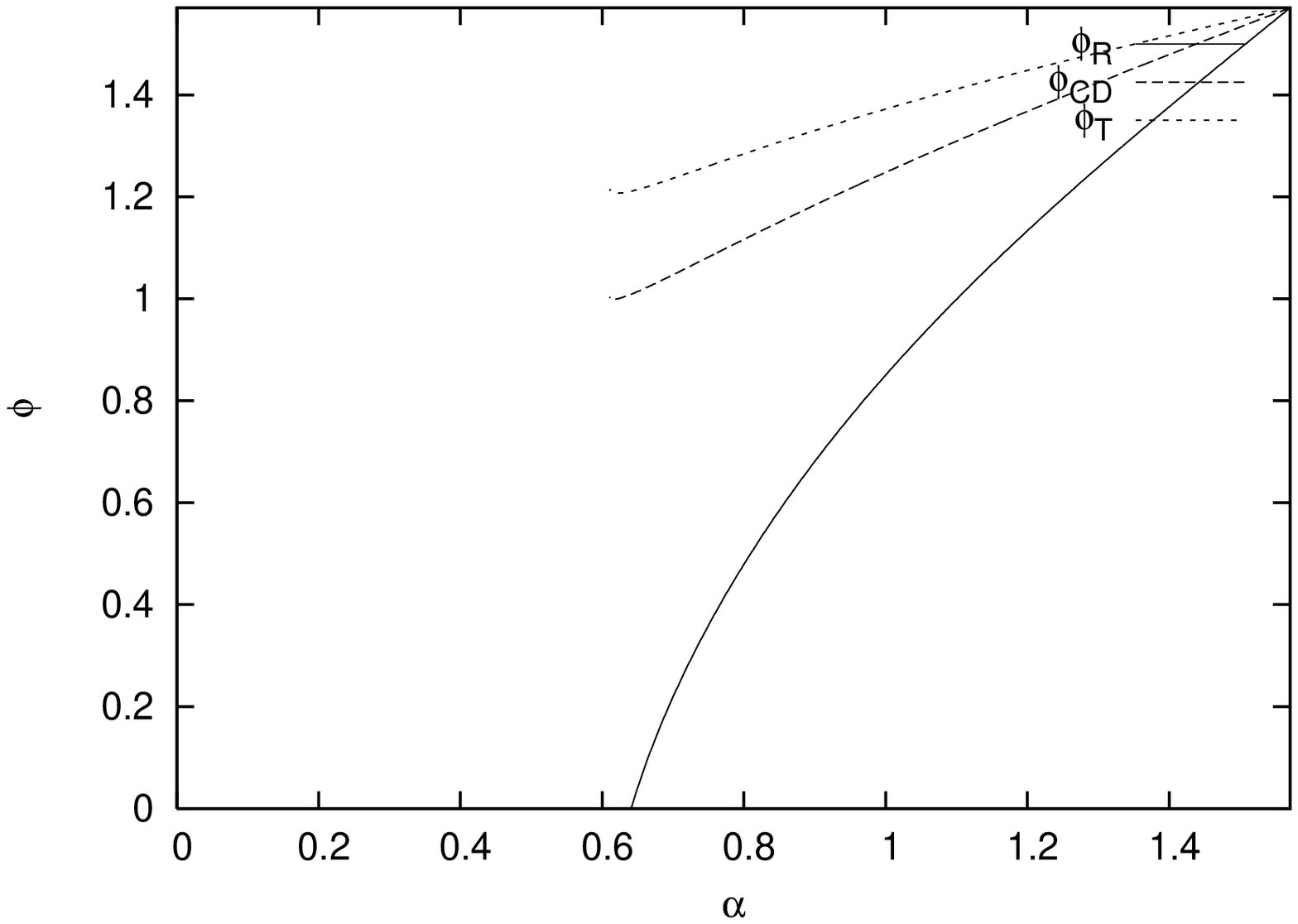}
\includegraphics[width=.32\textwidth,angle=-90]{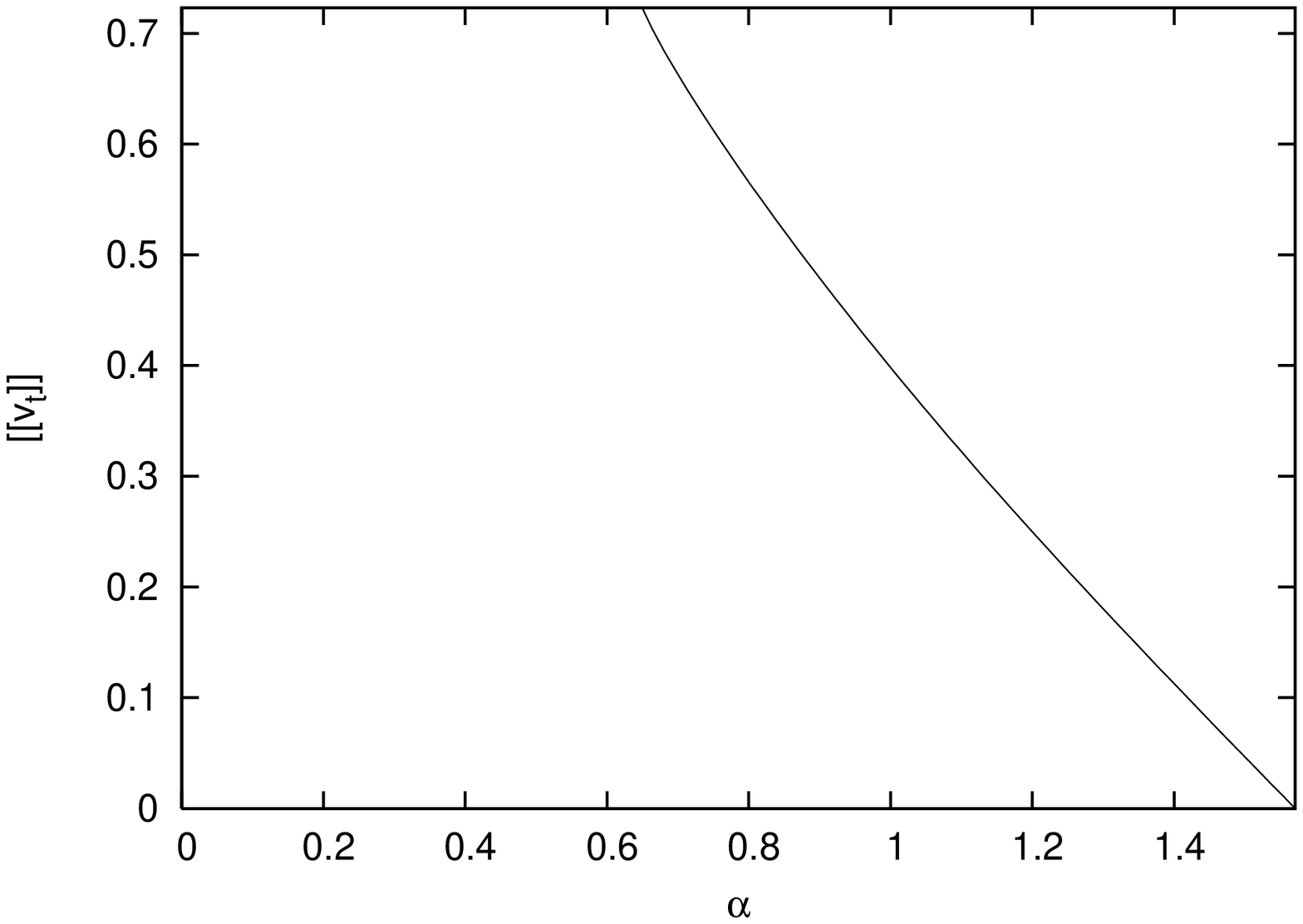}
\includegraphics[width=.32\textwidth,angle=-90]{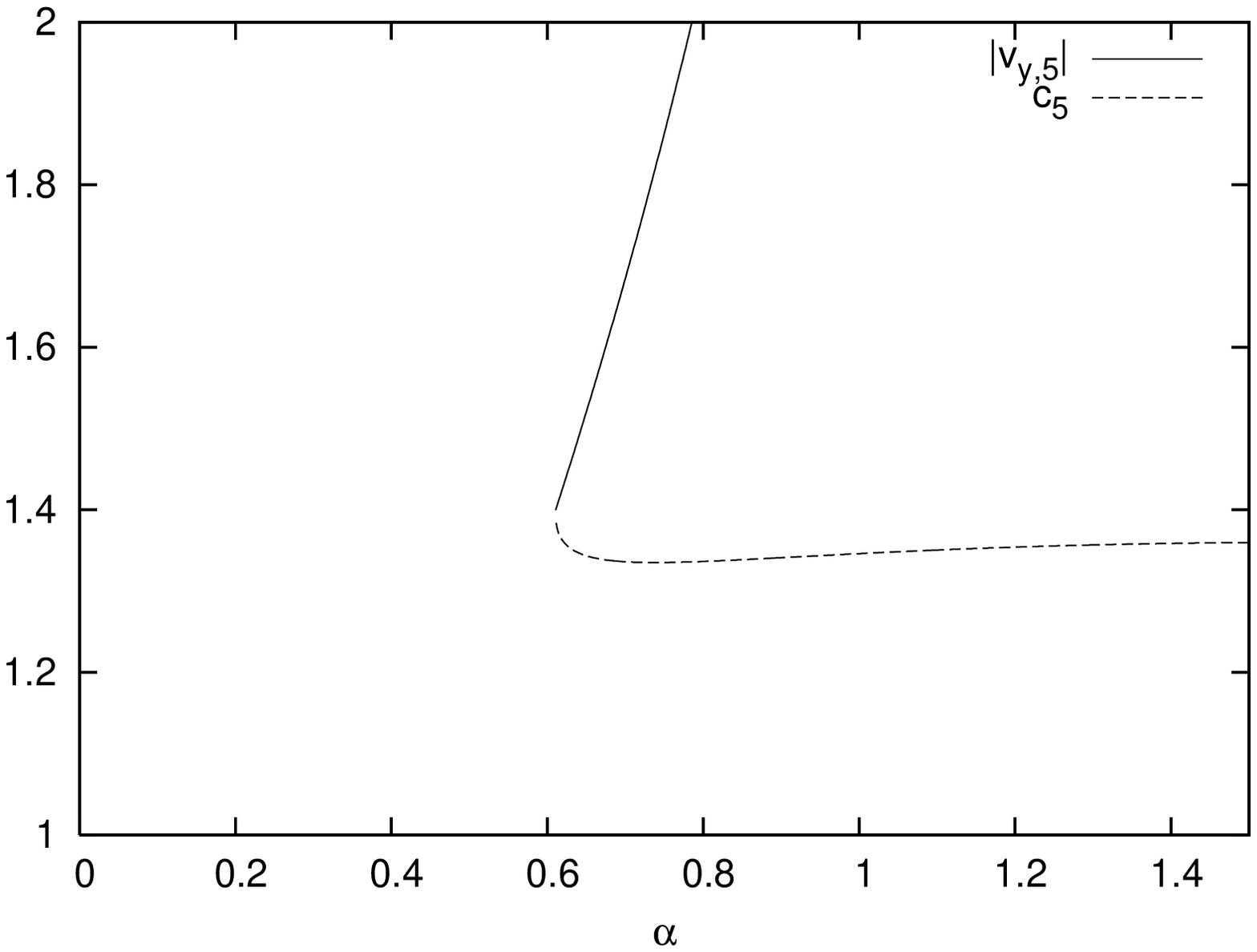}
\caption{\textit{Upper Left}: $p^*(\alpha)$.  Note that for $\alpha < 0.61$, there are no solutions for $p^*$: the refraction is irregular;    \textit{Upper Right}: the wave pattern for regular refraction; \textit{Lower Left}:  For $\alpha = \frac{\pi}{2}$, the problem is $1$-dimensional and there is no vorticity deposited on the interface.  For decreasing $\alpha$, the vorticity increases. \textit{Lower right:} For regular refraction, $|v_{y,5}| > \hat{c}_5$.}
\label{fig:alpha}
\end{figure}
Let us examine what the effect of the angle of incidence, $\alpha$, is.  Therefore we get back to the example from section $5.1$, $\left( \beta^{-1}, \gamma_l, \gamma_r, \eta, M \right) = \left(0, \frac{7}{5}, \frac{7}{5},3,2 \right)$ and let $\alpha$ vary: $\alpha \in \left]0, \frac{\pi}{2} \right]$.  Note that $\alpha = \frac{\pi}{2}$ corresponds to a $1$-dimensional Riemann problem.  The results are shown in figure~\ref{fig:alpha}.  Note that for regular refraction $v_{y,5}^2 > \hat{c}_5^2$.  We can understand this by noting that $\xi_{\pm}=\frac{v_{e,x} v_{e,y} \pm \hat{c}_{e} \sqrt{v_e^2 - \hat{c}_e^2}}{v_{e,x}^2 - \hat{c}_e^2} = \left(\frac{v_{e,x} v_{e,y} \mp \hat{c}_{e} \sqrt{v_e^2 - \hat{c}_e^2}}{v_{e,y}^2 - \hat{c}_e^2} \right)^{-1}= \hat{\xi}_{\mp}$, which are the eigenvalues of $\mathbf{G_u}^{-1}\cdot\mathbf{F_u} = (\mathbf{F_u}^{-1}\cdot\mathbf{G_u})^{-1}$.  Note that we could have started our theory from the quasilinear form $\mathbf{u}_y +(\mathbf{G_u}^{-1}\cdot\mathbf{F_u})\mathbf{u}_x = \mathbf{0}$ instead of equation (\ref{eq:quastat}).  If we would have done so, we would have found eigenvalues $\hat{\xi}$, which would correspond to $\frac{1}{atan \phi}$.  Moreover, both the eigenvalues, $\xi_+$ and $\xi_-$, have $4$ singularities, namely $\hat{c}_2 \in \{ -v_{x,2}, v_{x,2}, -v_{y,5}, v_{y,5}\}$ for $\xi_-$ and  $\hat{c}_5 \in \{ -v_{x,5}, v_{x,5}, -v_{y,2}, v_{y,2}\}$ for $\xi_+$, where thus $\hat{c}_5^2 = v_{5,y}^2 \Leftrightarrow \hat{c}_2^2 = v_2^2$ and $\hat{c}_2^2 = v_{y,2}^2 \Leftrightarrow \hat{c}_5^2 = v_5^2$.  It is now clear that it is one of the latter conditions that will be met for $\alpha_{crit}$.  In the example, the transition to irregular refraction occurs at $-v_{y,5}=\hat{c}_5$ and $
 \mathop {\lim }\limits_{\alpha \to \alpha_{crit}}p^*  = \frac{2 \gamma_r \eta M^2  tan^2 \left( \alpha_{crit} \right) - \gamma_l + 1}{\gamma_l + 1} = 6.67$.
\begin{figure}
\centering

\includegraphics[width=.99\columnwidth]{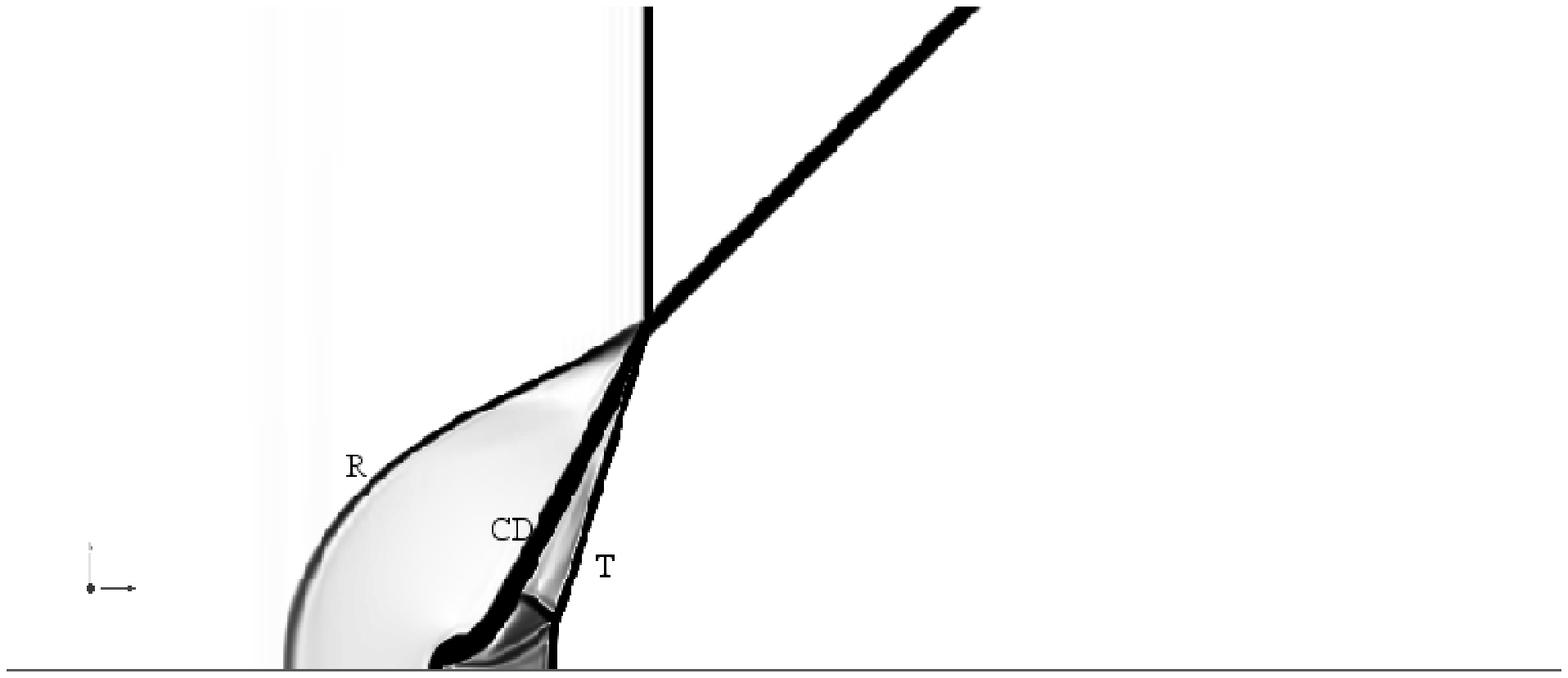}
\includegraphics[width=.99\columnwidth]{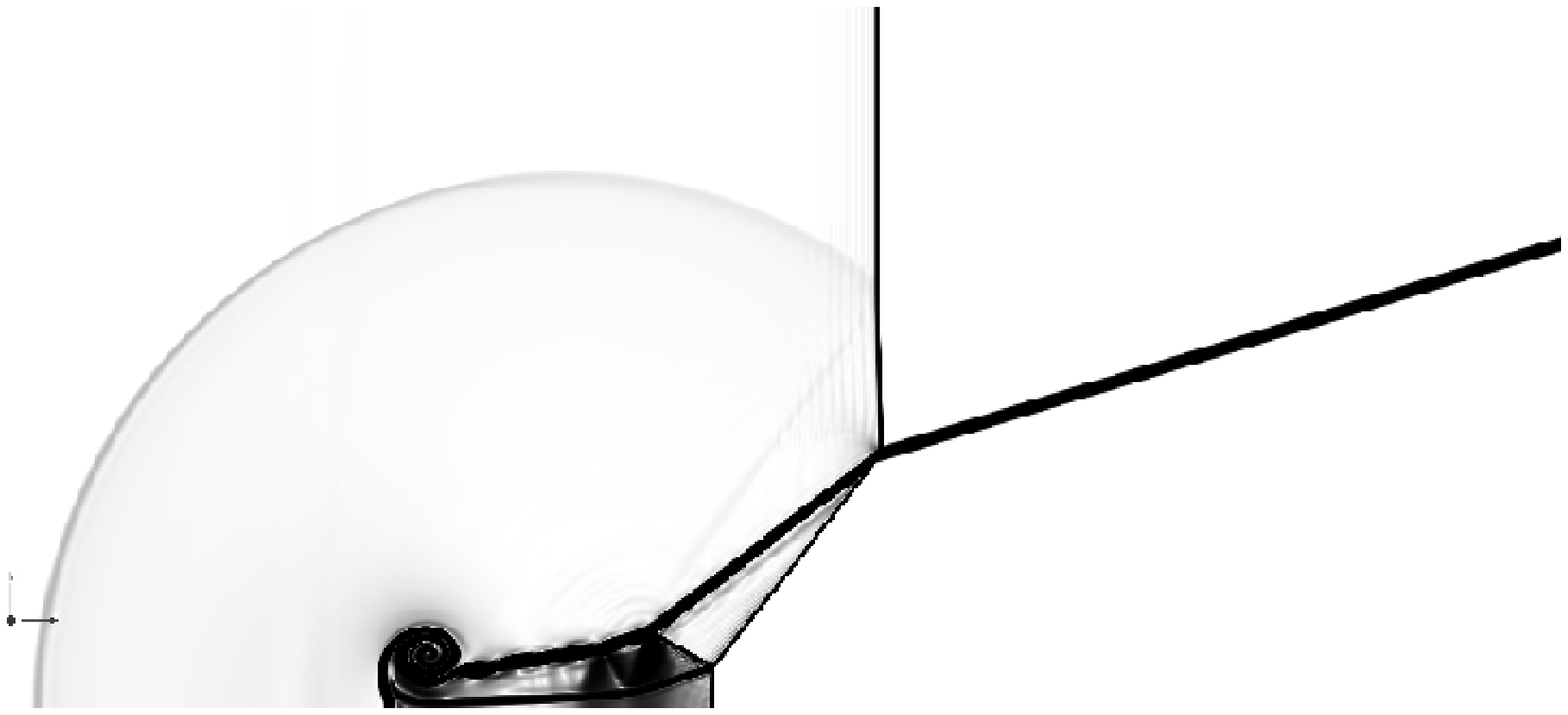}
\caption{ Schlieren plots of the density for $\left( \beta^{-1}, \gamma_l, \gamma_r, \eta, M \right) = \left( 0, \frac{7}{5}, \frac{7}{5},3,2 \right)$ with varying $\alpha$. \textit{Upper}: $\alpha = \frac{\pi}{4}$: a regular reference case.  \textit{Lower}: $\alpha = 0.3$: an irregular case.}

\label{fig:simalpha}
\end{figure}
Figure~\ref{fig:simalpha} shows Schlieren plots for density from AMRVAC simulations for the reference case $\alpha=\frac{\pi}{4}$, and the irregular case and $\alpha=0.3$.  In the regular case, all signals meet at the triple point, while for $\alpha < \alpha_{crit} = 0.61$, the signals do not meet at one triple point, the triple point forms a more complex structure and becomes irregular.  The CD, originated at the \textit{Mach stem}, reaches the triple point through an \textit{evanescent wave}, which is visible by the contourlines. This pattern is called \textit{Mach Reflection-Refraction}.  Decreasing $\alpha$ even more, the reflected wave transforms in a sequence of weak wavelets (see e.g. \cite{NO05}).  This pattern, of which the case $\alpha = 0.3$ is an example, is called \textit{Concave-Forwards irregular Refraction}.  These results are in agreement with our predictions.

\subsection{Abd-El-Fattah and Hendersons experiment}
\begin{figure}
\centering
\includegraphics[width=.32\textwidth,angle=-90]{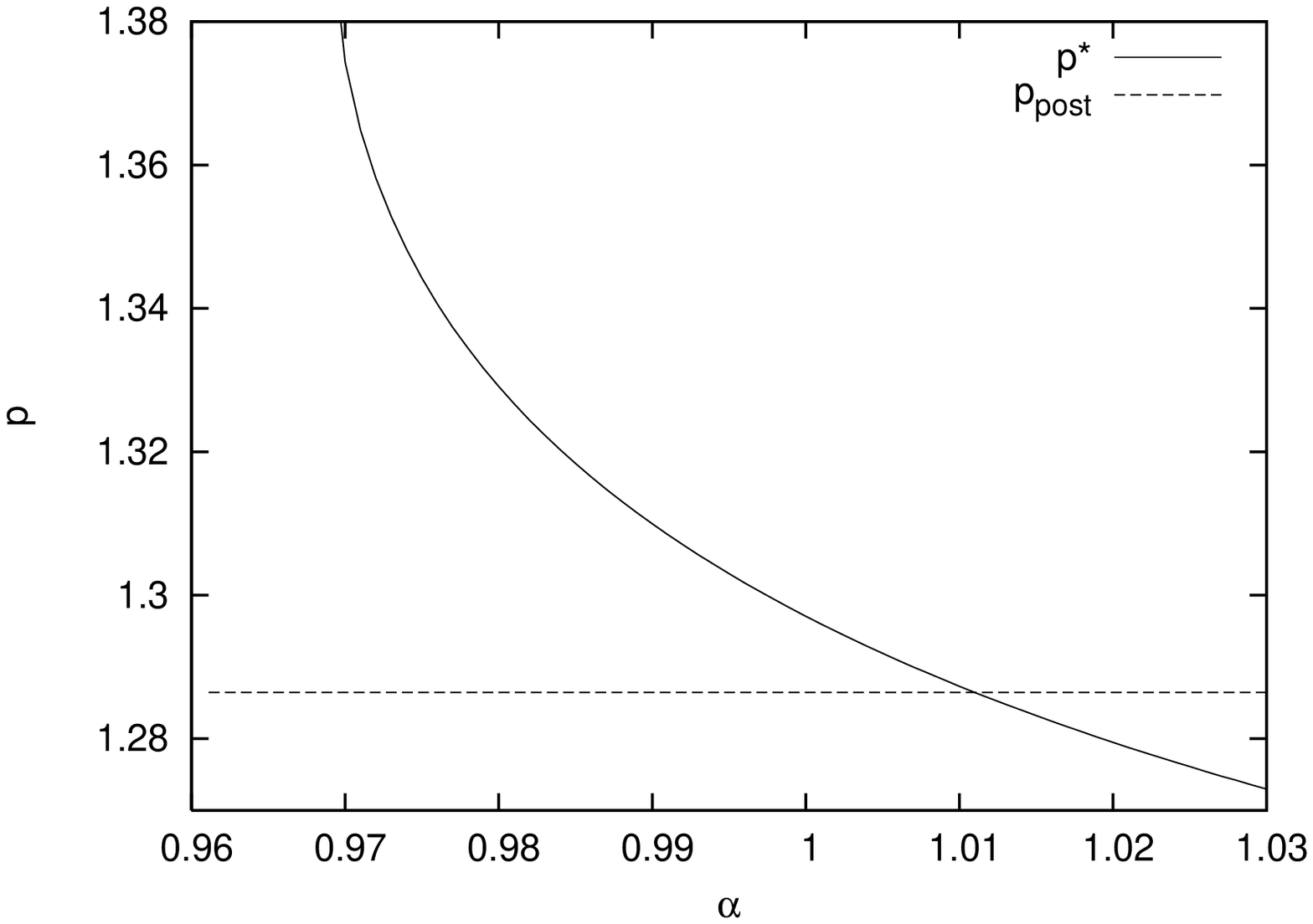}
\includegraphics[width=.32\textwidth,angle=-90]{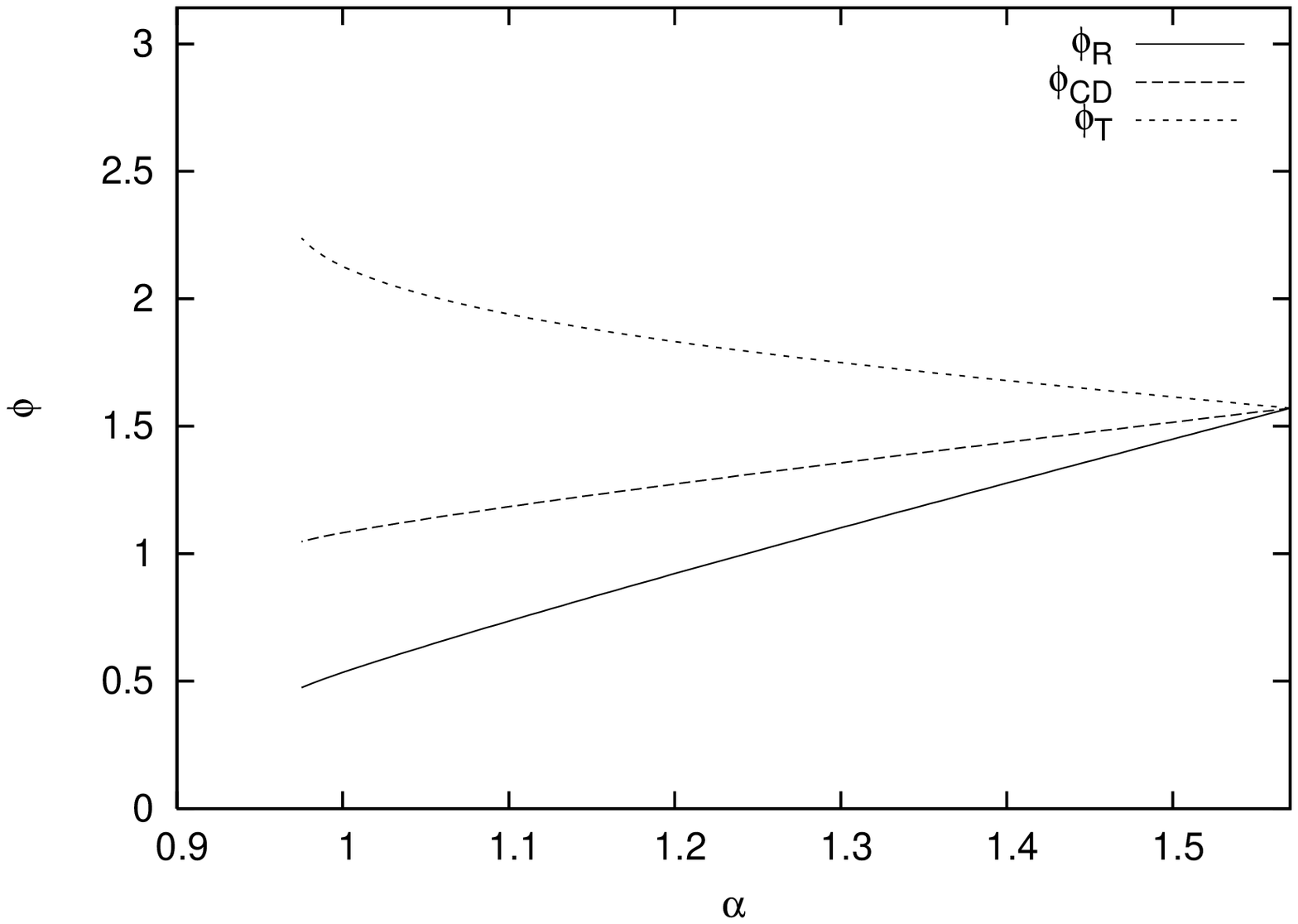}
\caption{Exact solution for the Abd-El-Fattah experiment.  \textit{Left}: $p^*(\alpha)$ confirms $\alpha_{crit} = 0.97$ and $\alpha_{trans}=1.01$. \textit{Right}: $\phi(\alpha)$.}
\label{fig:abdelfattah}
\end{figure}

In 1978, a shock tube experiment was performed by \cite{AH78b}. It became a typical test problem for simulations (see e.g. \cite{NO05}) and refraction theory (see e.g. \cite{HE91}). The experiment concerns a slow-fast shock refraction at a CO2/CH4 interface. The gas constants are $\gamma_{CO_2} = 1.288, \gamma_{CH_4} = 1.303, \mu_{CO_2} = 44.01$ and $\mu_{CH_4} = 16.04$.  Thus $\eta = \frac{\mu_{CH_4}}{\mu_{CO_2}} = 0.3645$.  A very weak shock, $M = 1.12$ is refracted at the interface under various angles.  \cite{NE43} theory predicts the critical angle $\alpha_{crit} = 0.97 $ and the transition angle $\alpha_{trans} = 1.01$, where the reflected signal is irregular if $\alpha < \alpha_{crit}$, a shock if $\alpha_{crit} < \alpha < \alpha_{trans}$ and an expansion fan if $\alpha_{trans} < \alpha$. This is in perfect agreement with the results of our solution strategy as illustrated in figure~\ref{fig:abdelfattah}.  There we show the pressure $p^*$ compared to the post shock pressure $p_{post}$, as well as the angles $\phi_R $, $\phi_{CD} $ and $\phi_T $ for varying angle of incidence $\alpha$.  Irregular refraction means that not all signals meet at a single point.  The transition at $\alpha_{crit}$ is one between a regular shock-shock pattern and an irregular \textit{Bound Precursor Refraction}, where the transmitted signal is ahead of the shocked contact and moves along the contact at nearly the same velocity.  This is also confirmed by AMRVAC simulations.  If the angle of incidence, $\alpha$, is decreased even further, the irregular pattern becomes a \textit{Free Precursor Refraction}, where the transmitted signal moves faster than the shocked contact, and reflects itself, introducing a side-wave, connecting $T$ to $CD$.  When decreasing $\alpha$ even further, another transition to the \textit{Free Precursor von Neumann Refraction} occurs.

\subsection{Connecting slow-fast to fast-slow refraction}

\begin{figure}
\includegraphics[width=.32\columnwidth,angle=-90]{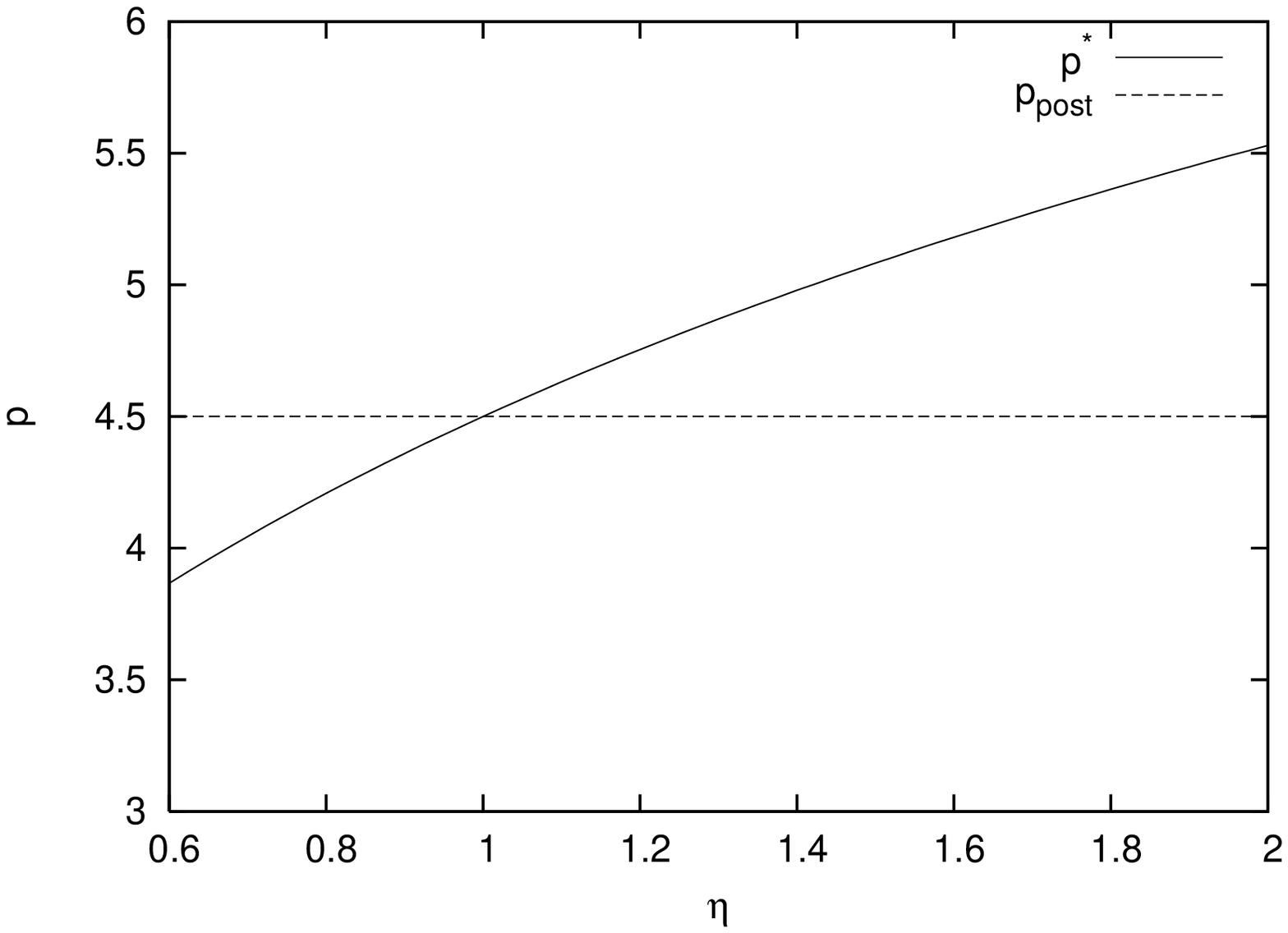}
\includegraphics[width=.32\columnwidth,angle=-90]{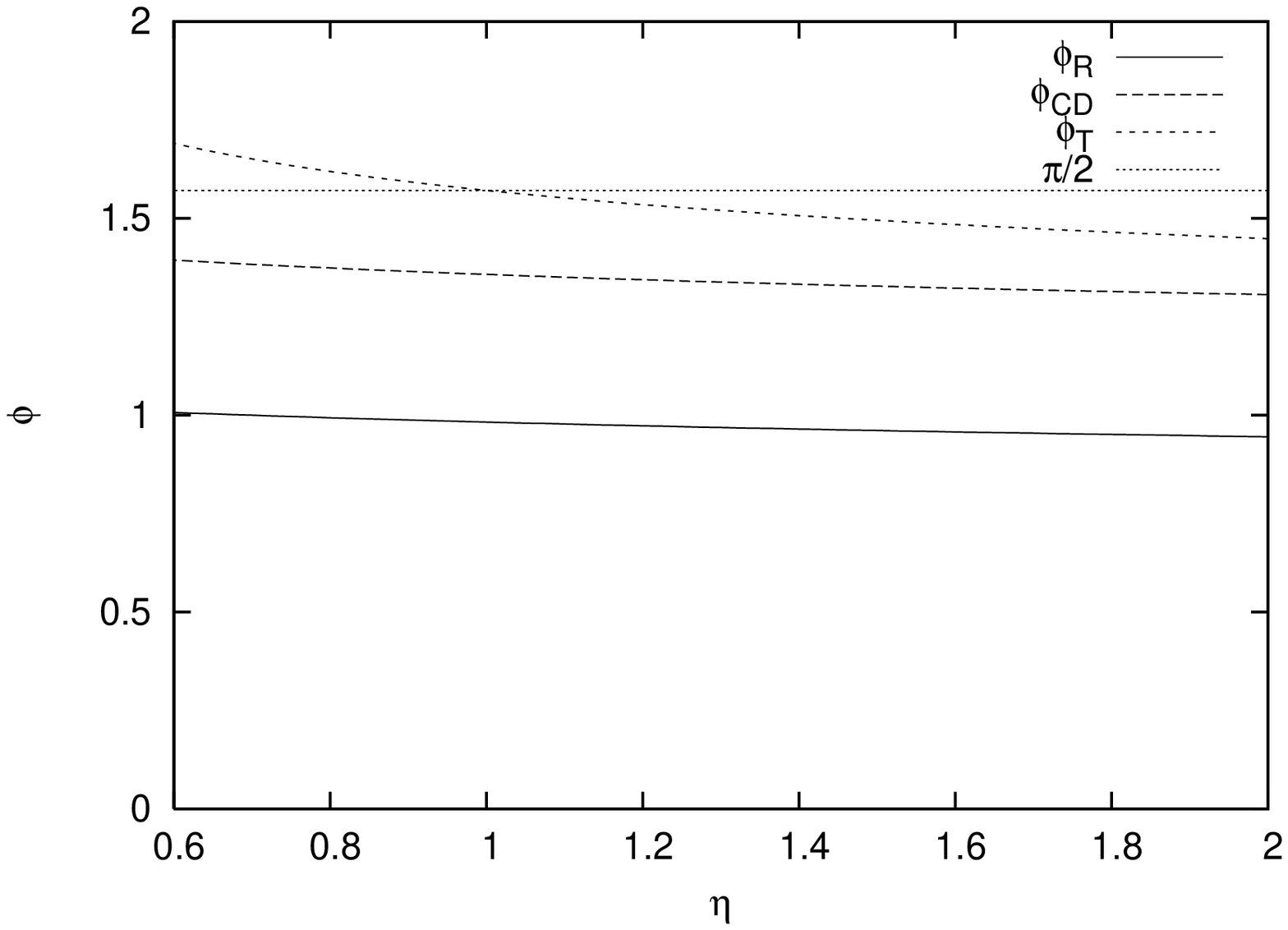}
\caption{Exact solution for $\left( \alpha, \beta^{-1}, \gamma_l, \gamma_r, M \right) =  \left( \frac{\pi}{4},0,\frac{7}{5},\frac{7}{5}, 2 \right)$ and a varying range of the density ratio $\eta$.    \textit{Left}: for $\eta<1$ we have $p^* < p_{post}=4.5$ and thus a reflected expansion fan, for $\eta>1$ we have $p^* > p_{post}=4.5$ and thus a reflected shock. \textit{Right}: for $\eta<1$: $\phi_T < \frac{\pi}{2}$ and for $\eta>1$: $\phi_T > \frac{\pi}{2}$.}
\label{fig:eta}
\end{figure}

Another example of how to trace transitions by the use of our solver is done by changing the density ratio $\eta$ across the CD.  Let us start from the example given in section $5.1$ and let us vary the value of $\eta$.

\begin{figure}
\centering
\includegraphics[width=.49\textwidth]{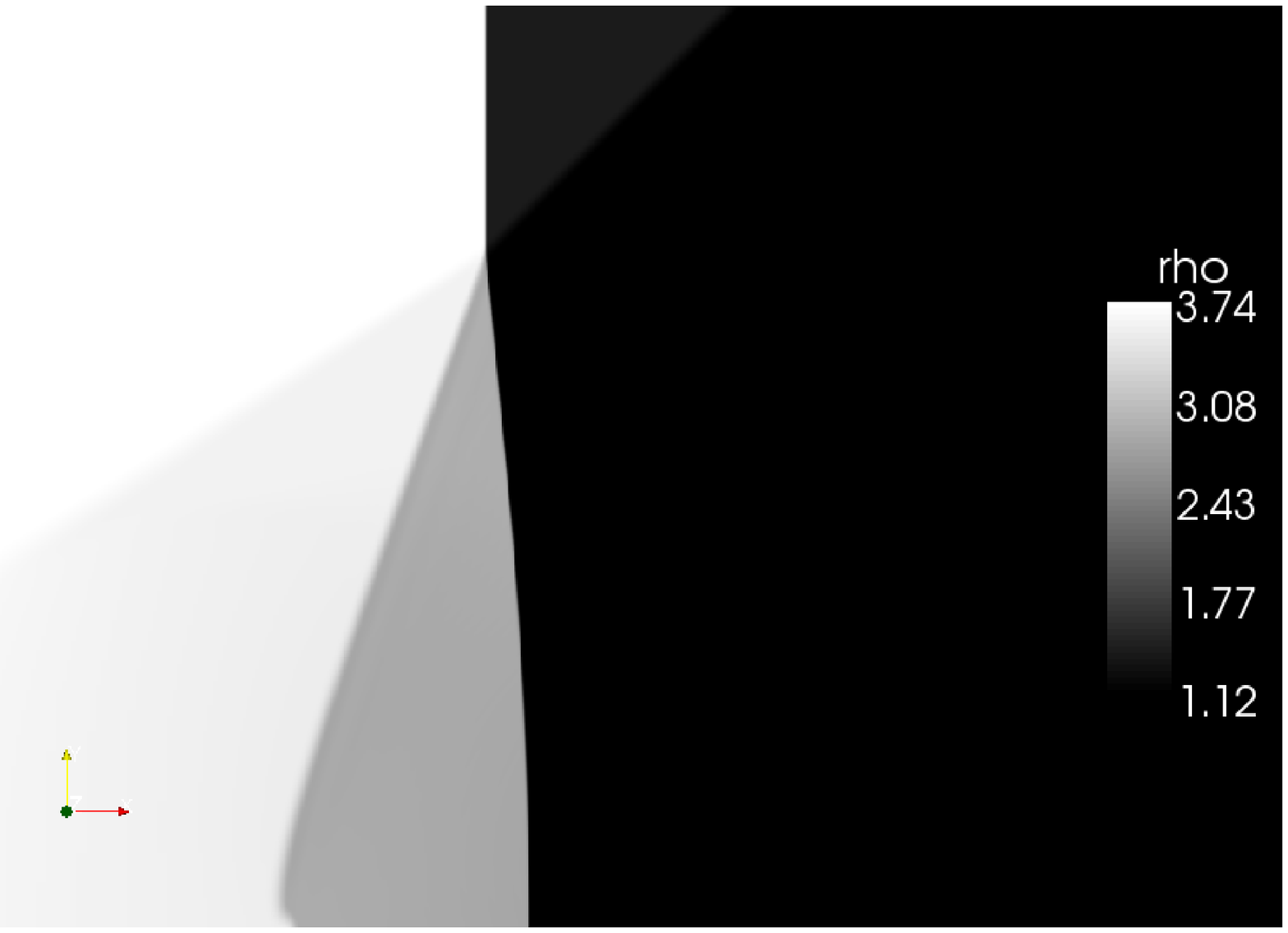}
\includegraphics[width=.49\textwidth]{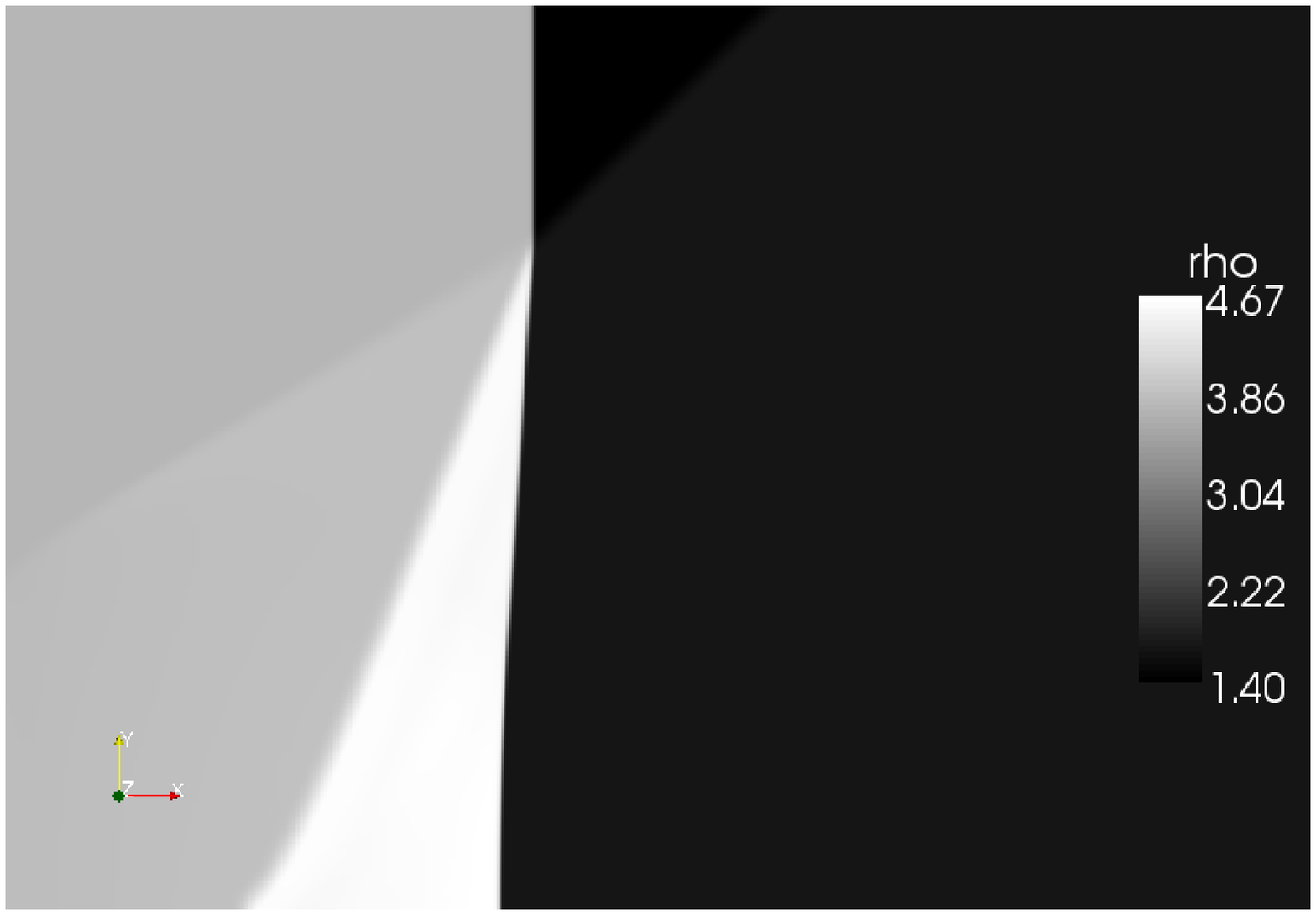}
\caption{ Density plots for $\left( \alpha, \beta^{-1}, \gamma_l, \gamma_r, M \right) = \left( \frac{\pi}{4}, 0, \frac{7}{5}, \frac{7}{5},2 \right)$. \textit{Left}: A slow/fast refraction with $\eta = 0.8$.  Note that $\phi_T > \frac{\pi}{2}$ and $R$ is an expansion fan.  \textit{Right}: A fast/slow refraction with $\eta = 1.2$.  Note that $\phi_T < \frac{\pi}{2}$ and $R$ is a shock.}
\label{fig:etasims}
\end{figure}

Here we have $\left( \alpha, \beta^{-1}, \gamma_l, \gamma_r, M \right) = \left( \frac{\pi}{4}, 0, \frac{7}{5}, \frac{7}{5},2 \right)$.  The results are shown in figure~\ref{fig:eta}.  Note that, since $p_{post}=4.5$, we have a reflected expansion fan for fast-slow refraction, and a reflected shock for slow-fast refraction.   The transmitted signal plays a crucial role in the nature of the reflected signal: for fast-slow refraction $\phi_T < \frac{\pi}{2}$, but for slow-fast refraction, $\phi_T > \frac{\pi}{2}$ and the transmitted signal bends forwards.  We ran our solver for varying values of $M$ and $\alpha$, and for all HD experiments with $\gamma_l = \gamma_r$, we came to the conclusion that a transition from fast-slow to slow-fast refraction, coincides with a transition from a reflected shock to a reflected expansion fan, with $\phi_T = \frac{\pi}{2}$.  This result agrees with AMRVAC simulations.  In figure~\ref{fig:etasims}, a density plot is shown for $\eta = 1.2$ and $\eta=0.8$.   

\subsection{Effect of a perpendicular magnetic field}

\begin{figure}
\centering

\includegraphics[width=.32\columnwidth,angle=-90]{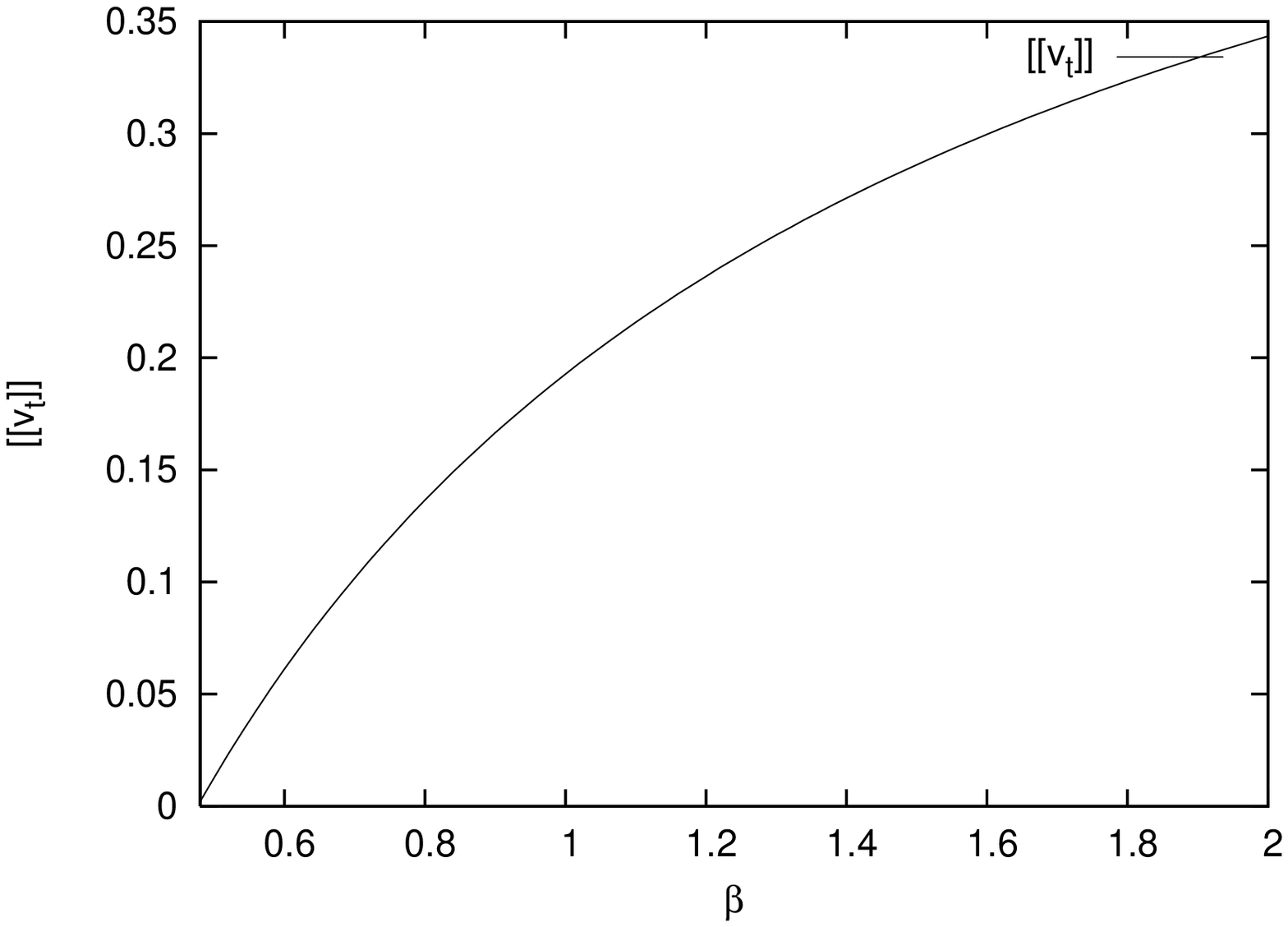}
\includegraphics[width=.32\columnwidth,angle=-90]{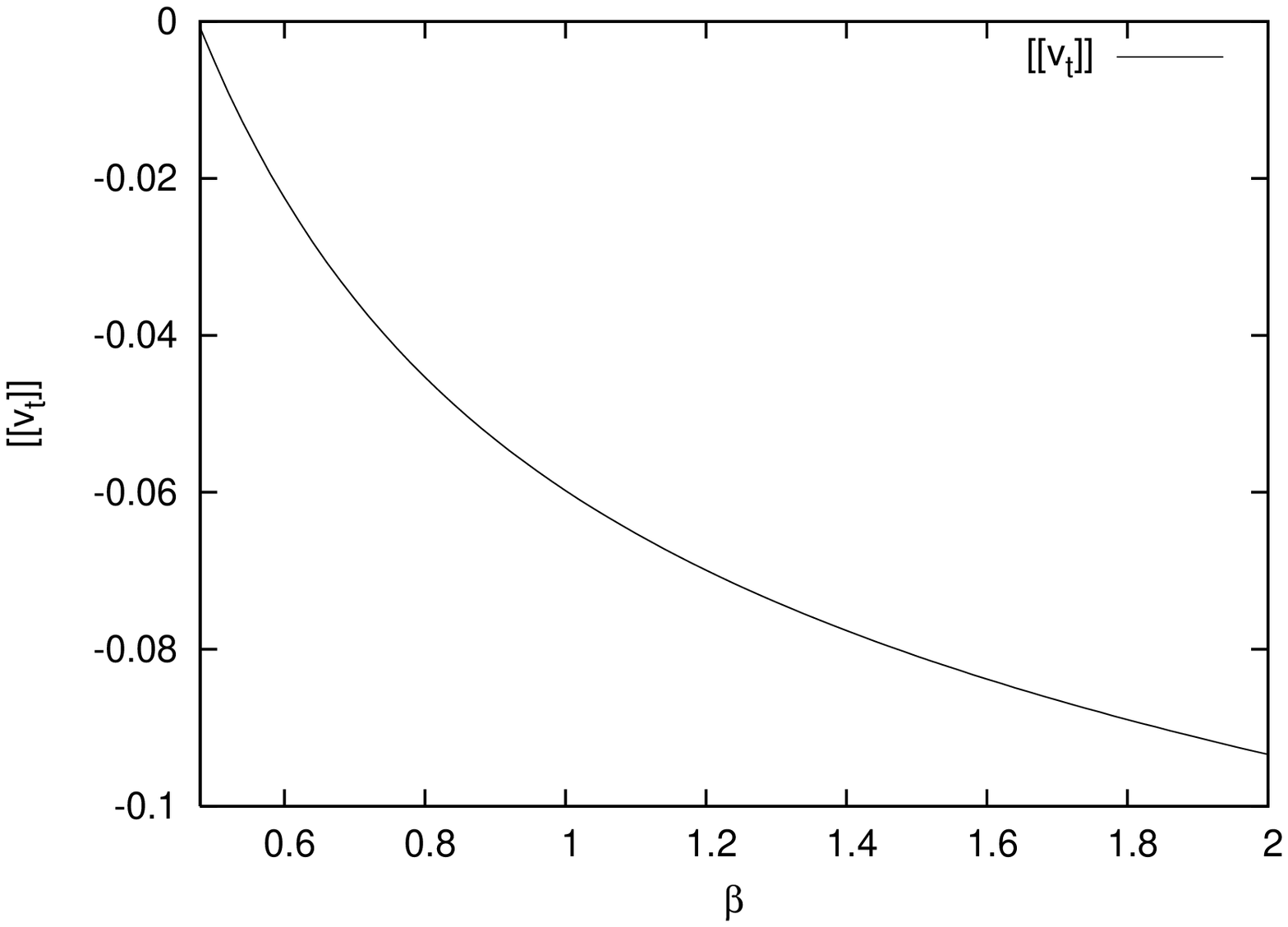}

\caption{ \textit{Left}: Solution for the fast-slow problem: strong perpendicular magnetic fields decrease the instability of the $CD$. \textit{Right}: Solution for the slow-fast problem: strong perpendicular magnetic fields decrease the instability of the $CD$.}
\label{fig:beffect}
\end{figure}

In general, the MHD equations result in the following jump conditions across a contact discontinuity
\begin{equation}
\left[ \left[ \begin{array}{c}
               p + \frac{B_t^2}{2}\\  B_n \\ B_n B_t \\ v_t B_n\\
              \end{array}
\right] \right] = \mathbf{0}.
\end{equation}
It follows, that if the component $B_n$ of the magnetic field, normal to the shock front is non-vanishing, a case we did not consider so far, the MHD equations do not allow for vorticity deposition on a contact discontinuity and the RMI is suppressed (\cite{WH05}).  The remaining question is what the effect of a purely tangential magnetic field is, where the field is perpendicular to the shock front and thus acts to increase the total pressure and the according flux terms.

Also note that it follows from equations 3.18 and 3.19 that $\frac{B}{\rho}$ is invariant across shocks and rarefaction fans.  Therefore, $\frac{B}{\rho}$ can only jump across the shocked and unshocked contact discontinuity and $B$ cannot change sign.
\begin{figure}
\centering
\includegraphics[width=.99\columnwidth]{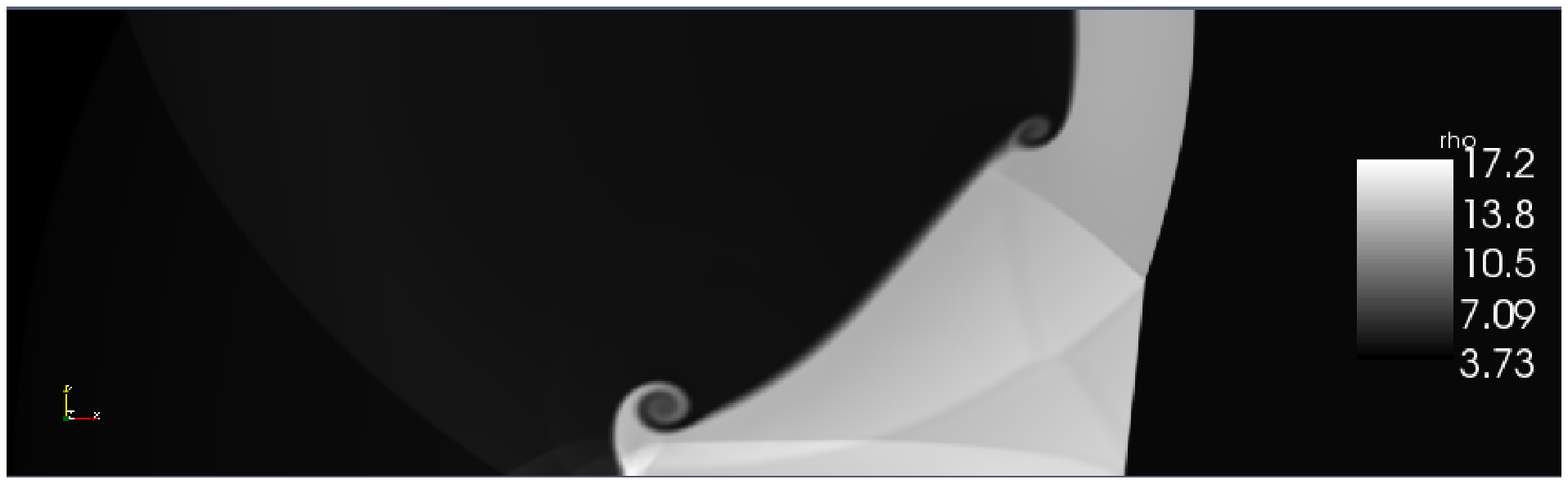}
\includegraphics[width=.99\columnwidth]{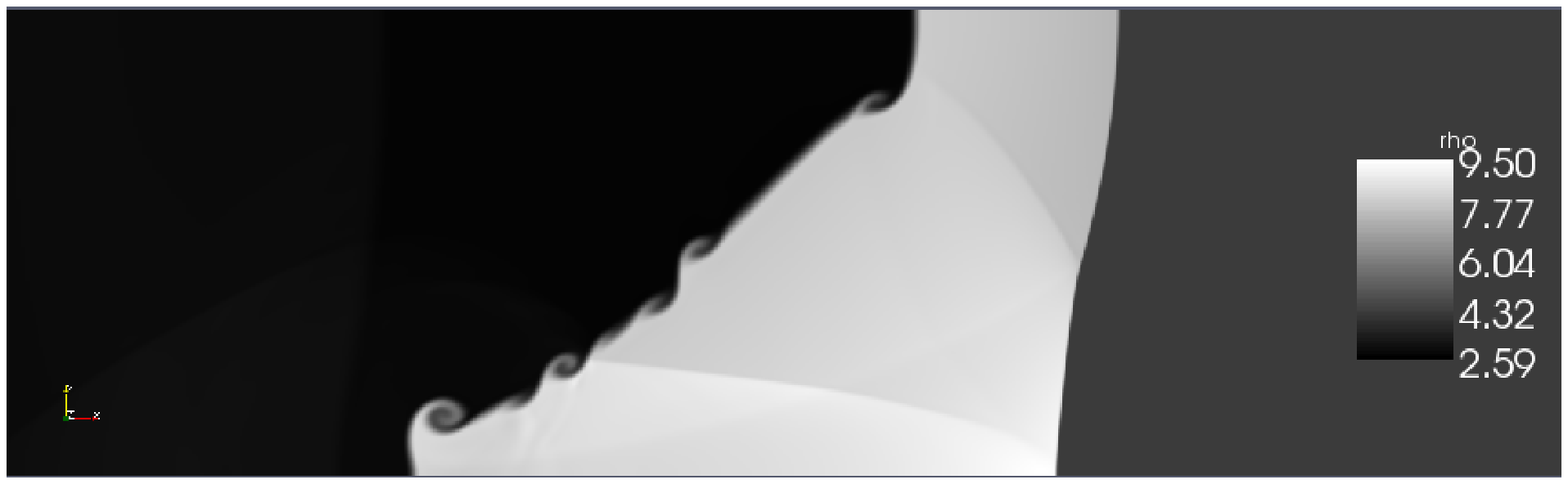}
\includegraphics[width=.99\columnwidth]{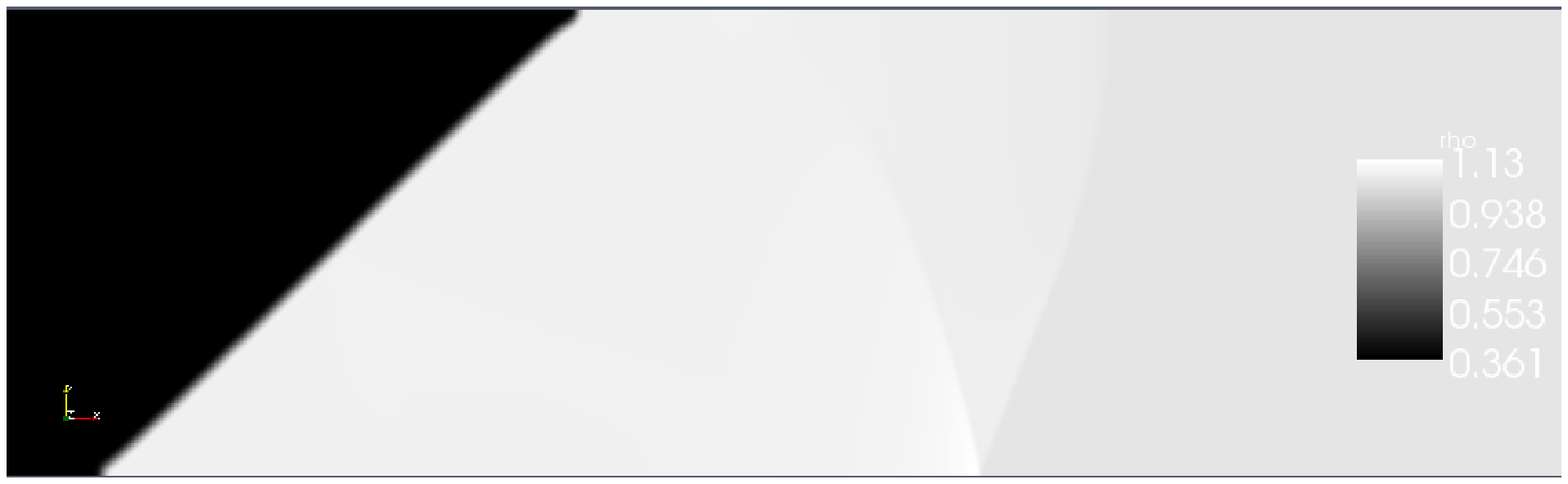}
\caption{ Density plots at $t=2.0$ for $\left( \alpha, \gamma_l, \gamma_r, \eta, M \right) = \left( \frac{\pi}{4}, \frac{7}{5}, \frac{7}{5},3,2 \right)$ with varying $\beta^{-1}$. \textit{Upper}: $\beta^{-1} = 0$.  The hydrodynamical Richtmyer-Meshkov instability causes the interface to roll up. \textit{Center}: $\beta^{-1} = \frac{1}{2}$.  Although the initial amount of vorticity deposited on the interface is smaller than in the HD case, the wall reflected signals pass the wall-vortex and interact with the $CD$, causing the $RMI$ to appear.  \textit{Lower}: $\beta^{-1} = 1$.  The shock is very weak and the interface remains stable.}
\label{fig:bsims}
\end{figure}

Revisiting the example from section $5.1$, we now let the magnetic field vary.  Figure~\ref{fig:beffect} shows $[[v_t]](\beta)$ across the CD.  Also for $\eta = 0.8$, making it a slow-fast problem,  $[[v_t]](\beta)$ is shown. First notice that no shocks are possible for $\beta < 0.476$, since $\omega_+$ would not satisfy $\omega_+ > -M$.  Manipulating equation~\ref{eq:omega}, we know that this is equivalent to 
\begin{equation}
 \beta > \beta_{min} \equiv \frac{2}{\gamma_l (M^2-1)}.
\label{eq:betamin}
\end{equation}
This relation is also equivalent to $c_1 > M$, which means that the shock is submagnetosonic, compared to the pre-shock region.  Figure \ref{fig:bsims} shows density plots from AMRVAC simulations at $t=2.0$, for $\left( \alpha, \gamma_l, \gamma_r, \eta, M \right) = \left( \frac{\pi}{4}, \frac{7}{5}, \frac{7}{5},3,2 \right)$ with varying $\beta^{-1}$.  First note that the interface is instable for the HD case.  Increasing $\beta^{-1}$ decreases the shock strength.  For $\beta^{-1}$ the interface remains stable, but for $\beta^{-1} = 1$, the shock is very weak: the Atwood number $\mathit{A}t = 0.17$, and the interface remains stable.

\begin{figure}
\centering
\includegraphics[width=.32\columnwidth,angle=-90]{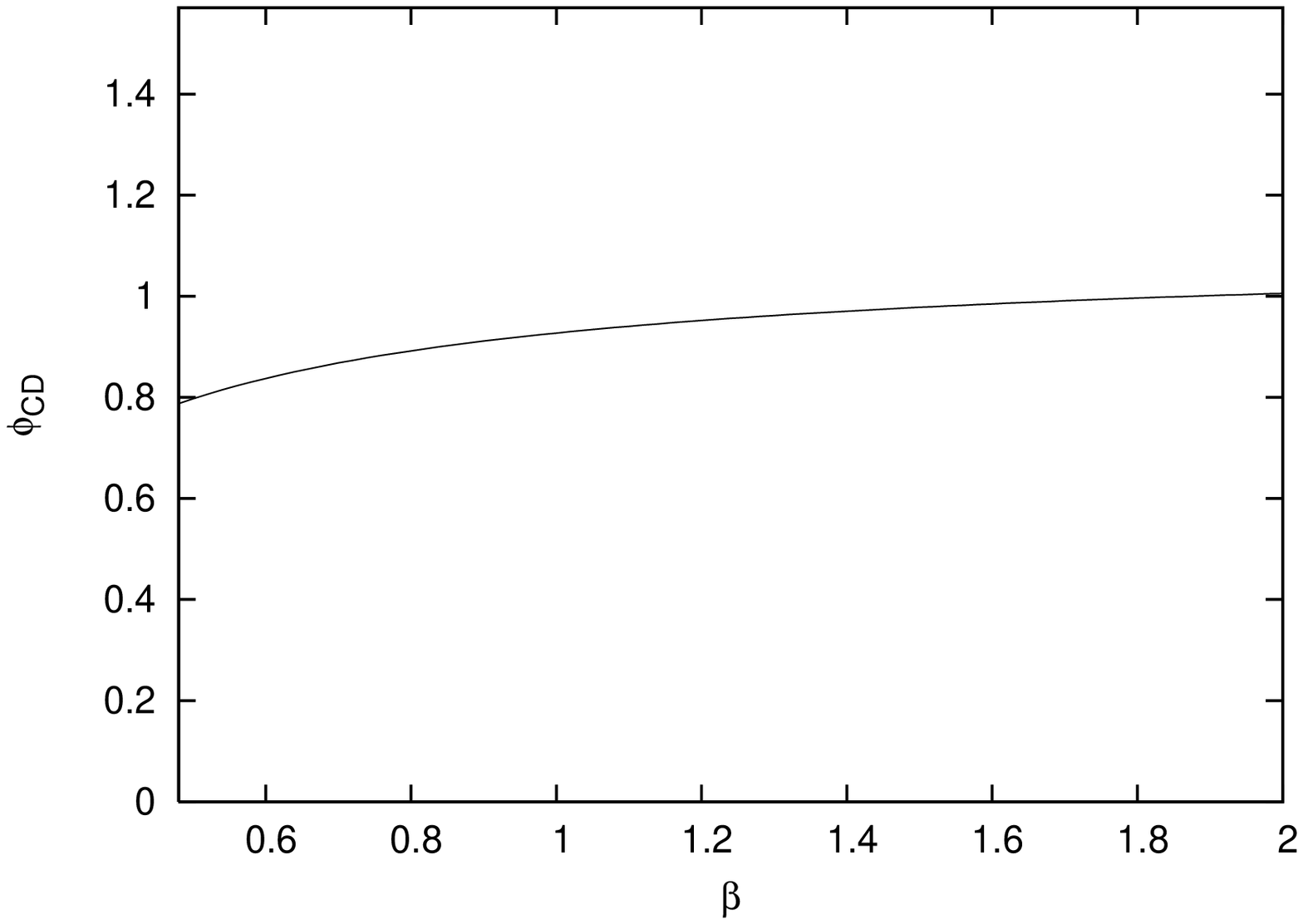}
\includegraphics[width=.32\columnwidth,angle=-90]{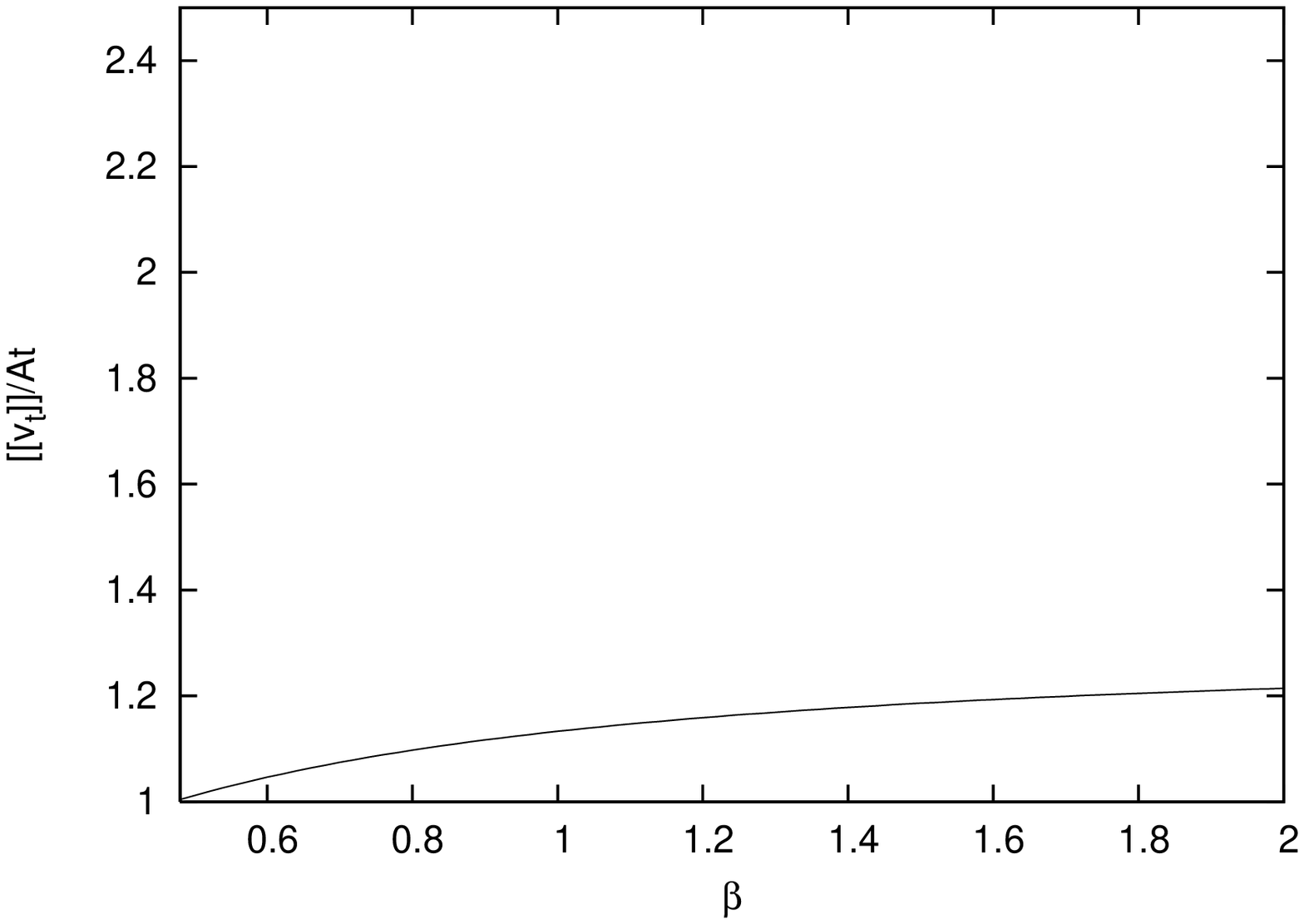}
\caption{The reference problem from \cite{SA03} with varying $\beta$. \textit{Left}: The dependence of $\phi_{CD}$ on $\beta$.  Note that  $\mathop {\lim }\limits_{\beta \to \beta_{min}} \phi_{CD} = \frac{\pi}{4}= \alpha$, since this is the limit to infinitely weak shocks: $\mathop {\lim }\limits_{\beta \to \beta_{min}} \mathit{A}t = 0$ \textit{Right}:  The vorticity deposition in the shocked contact scales as the Atwood number and $\mathop {\lim }\limits_{\beta \to \beta_{min}} \frac{[[v_t]]}{\mathit{A}t} = 1$. }
\label{fig:atwood}

\end{figure}

Shown in figure~\ref{fig:beffect}, is the vorticity across the CD.  In the limit case of this minimal plasma-$\beta$ the interface is stable, both for fast-slow and slow-fast refraction.  As expected, in the fast-slow case, the reflected signal is an expansion fan, while it is a shock in the fast-slow case.  Also note that the signs of the vorticity differ, causing the interface to roll up clockwise in the slow-fast regime, and counterclockwise in the fast-slow regime.  When decreasing the magnetic field, the vorticity on the interface increases in absolute value.  This can be understood by noticing that the limit case of minimal plasma-$\beta$ is also the limit case of very weak shocks.  This can for example be understood by noting that $ \mathop {\lim }\limits_{\beta \to \beta_{min}}  \phi_{CD} = \alpha$ (see figure~\ref{fig:atwood} ).  A convenient way to measure the strength of a shock is by use of its \textit{Atwood number}
\begin{equation} \mathit{A}t= \frac{\rho_2 - \rho_1}{\rho_2 + \rho_1}. \end{equation}
Figure~\ref{fig:atwood} shows the jump across the shocked contact $[[v_t]]$, scaled to the shocks Atwood number.  Note that in the limit case of very weak shocks the Atwood number equals the jump in tangential velocity across the $CD$, in dimensional notation:
\begin{equation} 
 \mathop {\lim }\limits_{\beta \to \beta_{min}}  \frac{\frac{[[v_t]]}{v_{s,1}}}{\mathit{A}t} = 1.
\end{equation}

\begin{figure}
\centering
\includegraphics[width=.32\columnwidth,angle=-90]{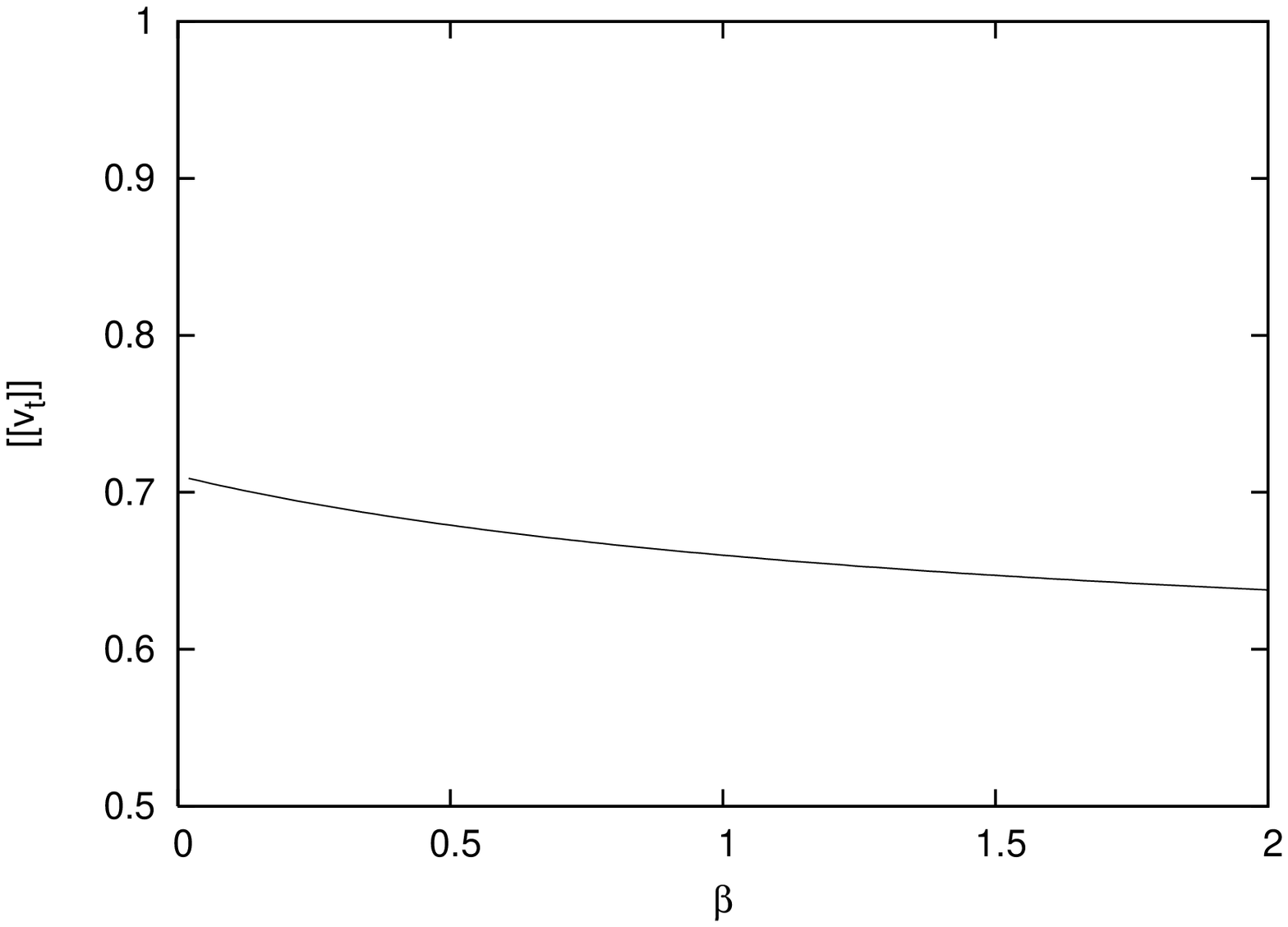}
\includegraphics[width=.32\columnwidth,angle=-90]{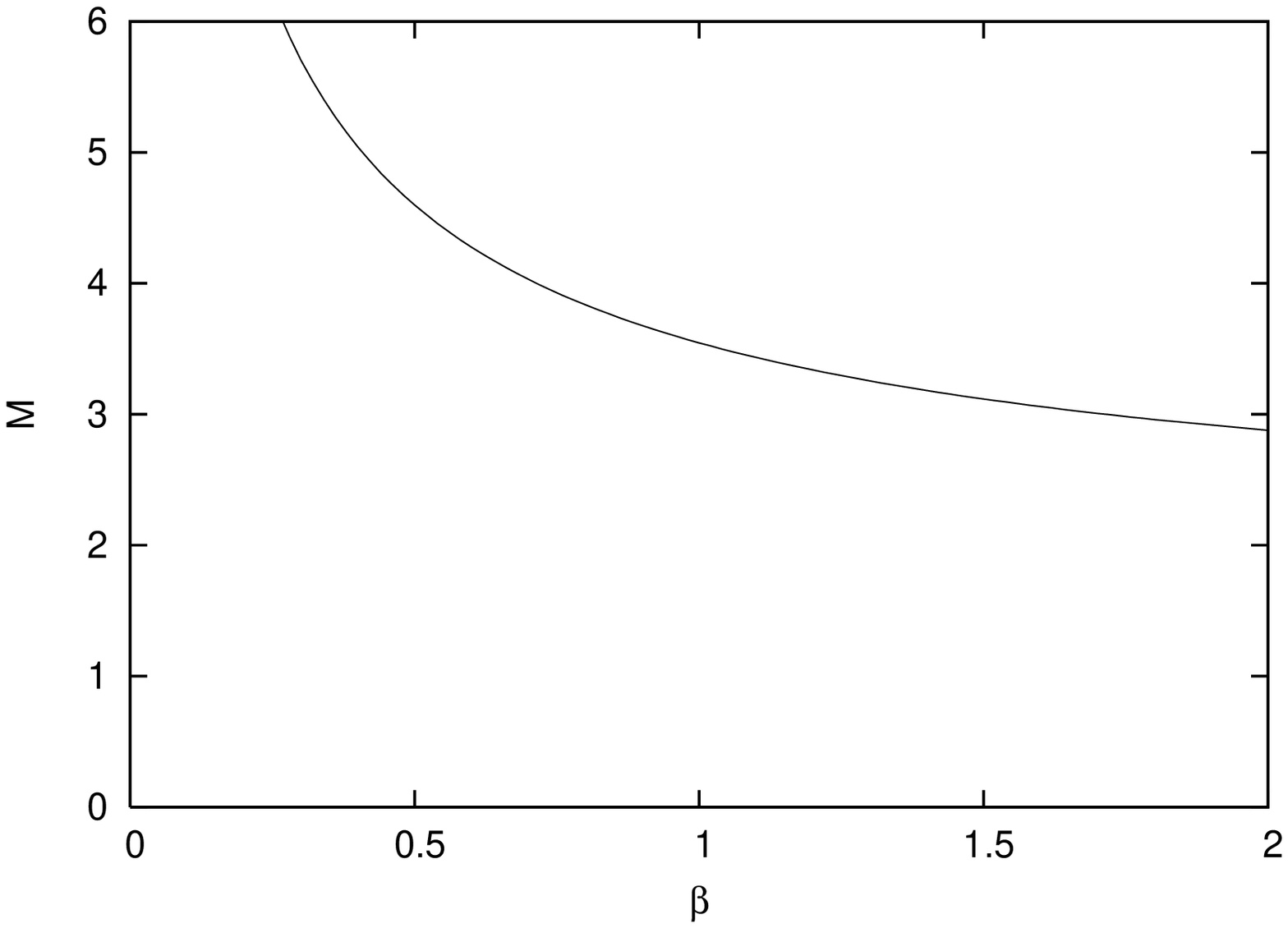}
\caption{ \textit{Left}: Solution for the fast-slow problem: strong perpendicular magnetic fields decrease the instability of the $CD$. \textit{Right}: Solution for the slow-fast problem: strong perpendicular magnetic fields decrease the instability of the $CD$.}
\label{fig:batwood}
\end{figure}

\begin{figure}
\centering
 \includegraphics[width=.99\columnwidth]{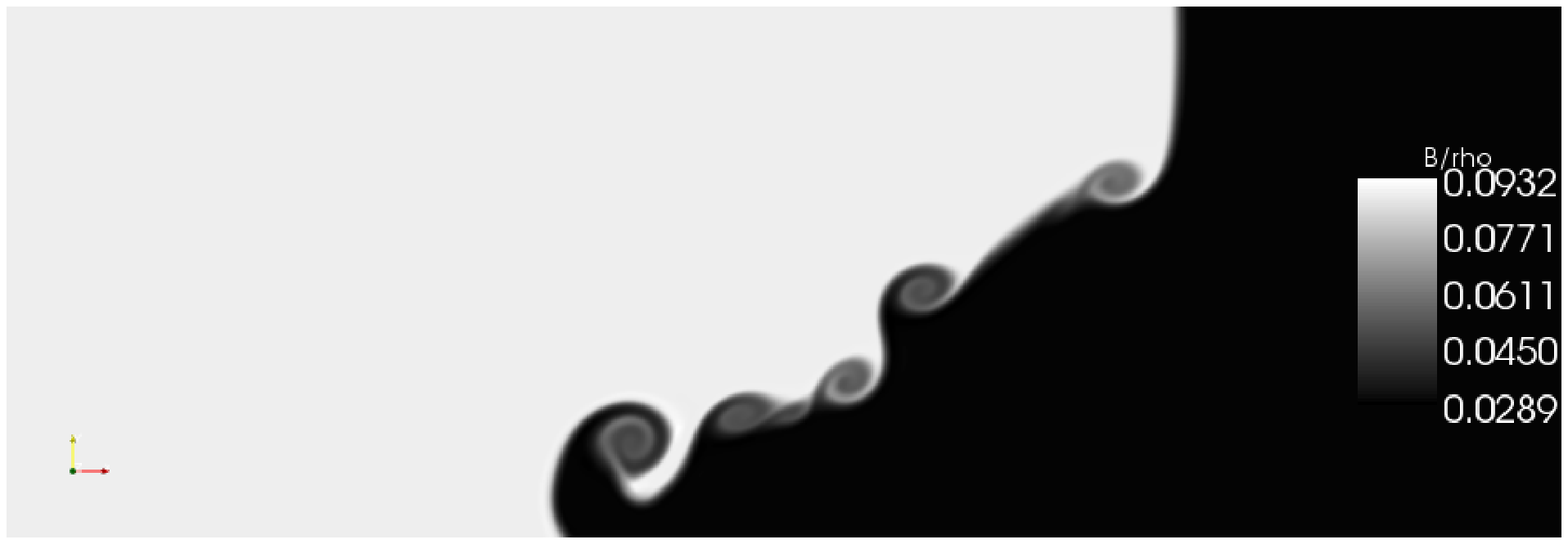}
 \includegraphics[width=.99\columnwidth]{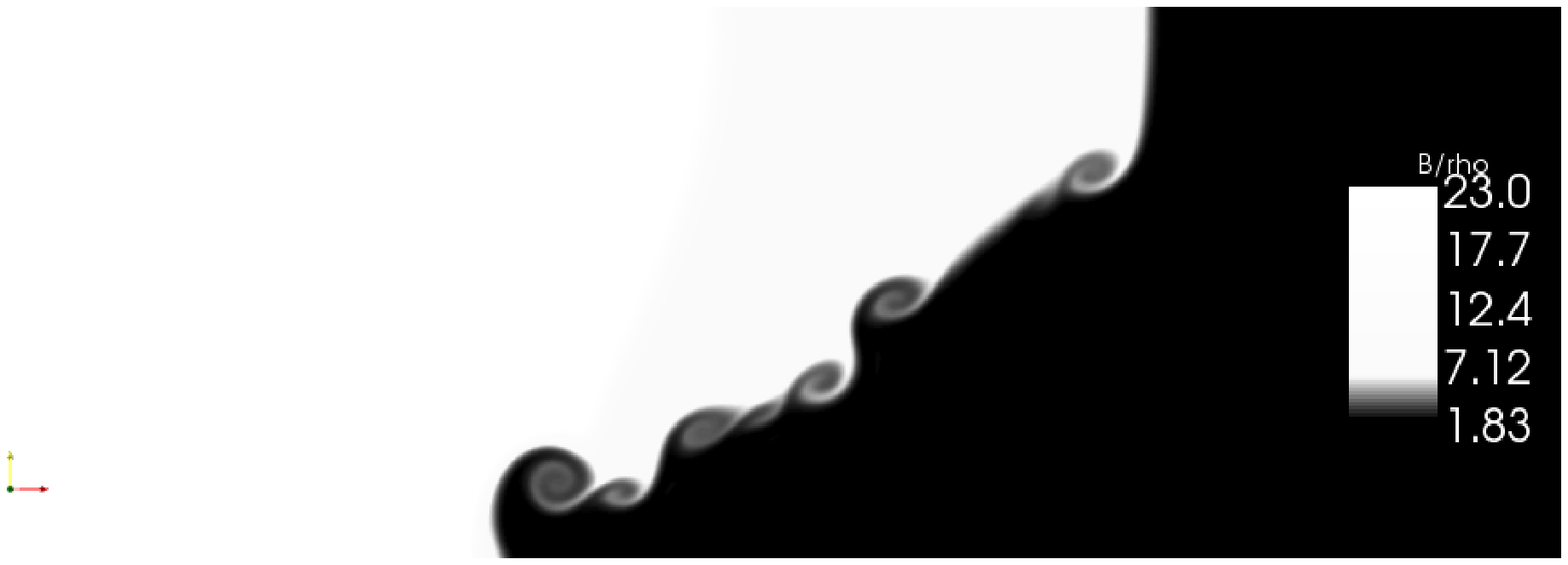}
\caption{ AMRVAC plots of $\frac{B}{\rho}$ for $\mathit{A}t = \frac{5}{11}$, with varying beta.  \textit{upper}: $\beta = 16$, \textit{lower}: $\beta= 0.25$.}
\label{fig:atwoodsims}
\end{figure}

When keeping the Atwood number constant, the shocks sonic Mach number is given by 
\begin{eqnarray}
 M &= &\frac{1+At}{1-At} \sqrt{\frac{(2- 2 \gamma - \gamma \beta)At^2 + (2 \gamma \beta + 2 \gamma )At - \gamma \beta -2}{(\gamma^2 \beta)At^2 + (\gamma^2 \beta -\gamma \beta)At -\gamma \beta}}\\
&=&\sqrt{\frac{(At+1)((\gamma \beta + 2 \gamma - 2)At - (\gamma \beta +2))}{\gamma \beta (1-At)(\gamma At + 1)}}.
\end{eqnarray}
Note that in the limit for weak shocks
\begin{equation} 
 \mathop {\lim }\limits_{At \to 0}  M = \sqrt{\frac{\gamma \beta +2}{\gamma \beta}},
\end{equation}
which is equivalent to \ref{eq:betamin}, and  in the limit for strong shocks, $M \rightarrow \infty$.  Figure \ref{fig:batwood} shows the deposition of vorticity on the shocked contact, for a constant Atwood number.  We conclude that under constant Atwood number, the effect of a perpendicular magnetic field is small: Stronger perpendicular magnetic field increase the deposition of vorticity on the shocked contact slightly.  This is confirmed by AMRVAC sumulations (see figure \ref{fig:atwoodsims}).

\section{Conclusions}

We developed an exact Riemann solver-based solution strategy for shock refraction at an inclined density discontinuity.  Our self-similar solutions agree with the early stages of nonlinear AMRVAC simulations.  We predict the critical angle $\alpha_{crit}$ for regular refraction, and the results fit with numerical and experimental results.  Our solution strategy is complementary to von Neumann theory, and can be used to predict full solutions of refraction experiments, and we have shown various transitions possible through specific parameter variations.  For perpendicular fields, the stability of the contact decreases slightly with decreasing $\beta$ under constant Atwood number.  We will generalise our results for arbitrary uniform magnetic fields, where up to $7$ signals arise.  In this case we will search for non-evolutionary solutions, involving intermediate shocks, and for alternative evolutionary solutions, where the appearance of intermediate shocks can be avoided by including compound waves.  We will investigate shock refraction involving initial slow, intermediate and fast shocks, and qualify the effect on the refraction.

\section{acknowlegdements}

The K.U.Leuven high performance computing cluster VIC has been used for all numerical simulations in this work.

\bibliographystyle{jfm}
\bibliography{jfm2esam}

\end{document}